\documentclass[a4paper,12pt]{article}
\usepackage{jheppub}

\title{Constraints on the second order transport coefficients of an uncharged fluid}

\author[a]{Sayantani Bhattacharyya}

\affiliation[a]{Harish Chandra Research Institute,\\
Chhatnag Rd, Jhunsi, Allahabad.}

\emailAdd{sayanta@hri.res.in}

\abstract{In this note we have tried to determine how the existence of
 a local entropy current with non-negative divergence 
constrains the second order transport
coefficients of an uncharged fluid, following the procedure described in
 \cite{Romatschke:2009kr}. Just on symmetry ground the stress tensor of 
an uncharged fluid 
can have 15 transport coefficients at second order in derivative expansion.
The condition of 
entropy-increase gives five relations among these 15 coefficients. 
So finally the relativistic stress tensor of an uncharged fluid can have 10 independent 
transport coefficients at second order. }

\begin{document}

\maketitle

\section{Introduction}\label{sec:intro}

Fluid dynamics is an effective description of near equilibrium physics. 
It captures the dynamics of locally equilibriated systems in which the 
parameters of equilibrium vary slowly compared to relaxation 
length scale. When, for instance,  microscopic dynamics is well described
by kinetic theory, the Boltzman equation reduces to the equations of fluid 
dynamics on length scales that are large compared to the molecular 
mean free path. 

The variables of fluid mechanics are the local values of the parameters 
that characterize fluid equilibrium; in the simplest context these are 
the fluid temperature, chemical potentials and velocity. The 
equations of fluid dynamics are simply the conservation of stress-tensor and 
all other charged currents, once the stress tensor and currents are 
expressed in terms of equilibriation parameters. The formulas that 
express the stress tensor and charge currents as functions of fluid 
variables are known as constitutive relations. 
As fluid dynamics is a long wavelength effective description, it is 
meaningful only to present constitutive relations in an expansion 
in derivatives of the fluid variables. 

For any given fluid, a microscopic 
computation of constitutive relations starting from a microscopic description of 
the system is often an impossibly difficult task. In the usual spirit 
of effective field theory, this task is, moreover, extraneous to the 
study of fluid dynamics. An autonomous `theory' of fluid dynamics 
addresses the following question: 
what is the most general form of the constitutive relations 
that could possibly arise in the fluid description of any consistent system 
(to any given order in the derivative expansion). 

The requirement of symmetry restricts the form of the constitutive relations. 
For example, the stress tensor is a tensor. At any given order in the 
derivative expansion, there exists only a finite number of (onshell inequivalent)
tensor structures one can build out of fluid variables and their derivatives. 
The most general expression for the stress tensor is clearly given by 
a linear combination of these inequivalent tensors, where the coefficients 
in this expansion are arbitrary functions of the scalar fluid variables 
(temperatures and chemical potentials in the simplest situations). While 
the requirements of symmetry are certainly necessary, they are not sufficient. 
There is atleast one additional constraint on allowed constitutive 
relations: that they are consistent with a local form of the second law 
of thermodynamics \cite{landau}. In other words any given constitutive relation must 
be accompanied by an entropy current, also constructed out of fluid 
variables. The entropy current must have the property that its divergence 
is positive for every conceivable fluid flow allowed by the fluid equations 
with the specified constitutive relations \footnote{However, the existence of 
a  local entropy current with positive divergence is really a heuristic, 
with as yet no solid basis
in thermodynamics or QFT}. 

It is a quite remarkable fact that the requirement of the existence of 
a positive divergence entropy current constrains the allowed constitutive
relations of fluid dynamics in a quite dramatic manner; as we will see 
in some detail below, this requirement reduces the number of free parameters
(or more precisely free functions) allowed in constitutive relations. 

In this note we will work the restrictions imposed by this requirment 
on the constitutive relations of an uncharged relativistic fluid in $3+1$ 
spacetime dimensions at second 
order in the derivative expansion. In this system the variables of fluid 
dynamics are simply the temperature $T$ and the fluid four velocity $u^\mu$. 
As we will see in some detail below, symmetry considerations allow a 15 
parameter worth of constitutive relations  for the stress tensor, where 
every parameter is an arbitrary function of the temperature. It was already 
noted by Romatschke \cite{Romatschke:2009kr} that the requirement of 
entropy increase 
imposes atleast two relations between these 15 functions. In this note 
we generalize Romatschke's analysis and demonstrate that a complete study 
of the requirements of positivity of the entropy current imposes 5 relations 
on the 15 coefficients described above. The set of all 2nd order constitutive
relations consistent with the positivity of entropy increase 
is parameterized by ten functions of the temperature. But unlike the first order
case we did not find any inequalities among the second order transport coefficients.

We now give a detailed presentation of our final results, i.e. 
a precise characterization of the 10 parameter set of allowed constitutive
relations at second order in the derivative expansion for uncharged 
relativistic fluids. Let the fluid stress tensory be given by 
$$T_{\mu\nu}= T^{perf}_{\mu\nu} + \Pi^{\mu\nu}$$
We work in the so called Landau frame which imposes the transversality 
condition 
$$ u^\mu \Pi_{\mu\nu}=0 $$
In this frame the most general allowed form for $\Pi_{\mu\nu}$ upto 
second order in the derivative expansion is given by\footnote{Our convention for
Riemann tensor is the following
$${R^\rho}_{\alpha\beta\nu} = \partial_\beta \Gamma^\rho_{\alpha\nu} -
\partial_\nu \Gamma^\rho_{\alpha\beta} 
+ \Gamma^\lambda_{\alpha\nu}\Gamma^\rho_{\lambda\beta}
-\Gamma^\lambda_{\alpha\beta}\Gamma^\rho_{\lambda\nu}$$}

\begin{equation}\label{stresderi1}
\begin{split}
\Pi_{\mu\nu} =~&-\eta\sigma_{\mu\nu} - \zeta P_{\mu\nu} \Theta\\
~&T\bigg[ \tau ~(u.\nabla)\sigma_{\langle\mu\nu\rangle} + \kappa_1 R_{\langle \mu\nu\rangle} + \kappa_2 F_{\langle \mu\nu\rangle} +\lambda_0~ \Theta\sigma_{\mu\nu}\\
&+ \lambda_1~ {\sigma_{\langle \mu}}^a\sigma_{a\nu\rangle}+ \lambda_2~ {\sigma_{\langle \mu}}^a\omega_{a\nu\rangle}+ \lambda_3~ {\omega_{\langle \mu}}^a\omega_{a\nu\rangle} + \lambda_4~{\mathfrak a}_{\langle\mu}{\mathfrak a}_{\nu\rangle}\bigg]\\
&+TP_{\mu\nu}\bigg[\zeta_1(u.\nabla)\Theta + \zeta_2 R + \zeta_3R_{00}
 + \xi_1 \Theta^2 + \xi_2 \sigma^2+ \xi_3 \omega^2 
+\xi_4 {\mathfrak a}^2 \bigg]
\end{split}
\end{equation}
where 
\begin{equation}\label{notation}
\begin{split}
&u^\mu =\text{The normalised four velocity of the fluid}\\
&P^{\mu\nu} = g^{\mu\nu} + u^\mu u^\nu =\text{Projector perpendicular to $u^\mu$}\\
&\Theta = \nabla. u = \text{Expansion},~~
{\mathfrak a}_\mu = (u.\nabla) u_\mu = \text{Acceleration}\\
&\sigma^{\mu\nu} = 
P^{\mu\alpha} P^{\nu\beta}\left(\frac{\nabla_\alpha u_\beta + \nabla_\beta u_\alpha}{2}
 - \frac{\Theta}{3}g_{\alpha_\beta}\right) = \text{Shear tensor}\\
&\omega^{\mu\nu} = 
P^{\mu\alpha} P^{\nu\beta}\left(\frac{\nabla_\alpha u_\beta 
- \nabla_\beta u_\alpha}{2}\right)=\text{Vorticity}\\
&F^{\mu\nu} = R^{\mu a \nu b}u_a u_b,~~R^{\mu\nu} 
= R^{a\mu b\nu}g_{ab}~~(R^{abcd} = \text{Reimann tensor})\\
&\sigma^2 = \sigma_{\mu\nu}\sigma^{\mu\nu},~~~~\omega^2 = \omega_{\mu\nu}\omega^{\nu\mu}
\end{split}
\end{equation}
and
$$
A_{\langle\mu\nu\rangle} \equiv P_\mu^\alpha P_\nu^\beta\left(\frac{A_{\alpha\beta} + A_{\beta\alpha}}{2} - \left[\frac{A_{ab}P^{ab}}{3}\right]g_{\alpha\beta}\right)~~\text{For any tensor $A_{\mu\nu}$}
$$

It turns out that `entropy-positivity' does not impose any constraint 
on $\tau,~\lambda_0,~\lambda_1,~\lambda_2,~\zeta_1,~\xi_1$ and $\xi_2$.
The rest of the eight second order transport coefficients satisfy the following 
5 relations. 
\begin{equation}\label{relationsintro}
  \begin{split}
  \kappa_2 =&~ \kappa_1 + T\frac{d\kappa_1}{dT}\\
  \zeta_2 =&~ \frac{1}{2}\left[s\frac{d\kappa_1}{ds} - \frac{\kappa_1}{3}\right]\\
  \zeta_3 = &\left(s\frac{d\kappa_1}{ds} + \frac{\kappa_1}{3}\right) + \left(s\frac{d\kappa_2}{ds} - \frac{2\kappa_2}{3}\right)+\frac{s}{T}\left(\frac{dT}{ds}\right)\lambda_4\\
  \\
  \xi_3=&~\frac{3}{4}\left(\frac{s}{T}\right)\left(\frac{dT}{ds}\right)\left(T\frac{d\kappa_2}{dT} + 2\kappa_2\right) -\frac{3\kappa_2}{4} +\left(\frac{s}{T}\right)\left(\frac{dT}{ds}\right)\lambda_4 \\
  &+\frac{1}{4}\left[s\frac{d\lambda_3}{ds} + \frac{\lambda_3}{3} -2 \left(\frac{s}{T}\right)\left(\frac{dT}{ds}\right)\lambda_3\right]\\
  \xi_4 =&~-\frac{\lambda_4}{6} - \frac{s}{T}\left(\frac{dT}{ds}\right)\left(\lambda_4 + \frac{T}{2}\frac{d\lambda_4}{dT}\right) 
  -T\left(\frac{d\kappa_2}{dT}\right)\left(\frac{3s}{2T}\frac{dT}{ds} - \frac{1}{2}\right) \\
  &- \frac{Ts}{2} \left(\frac{dT}{ds}\right)\left(\frac{d^2\kappa_2}{dT^2}\right)
  \end{split}
  \end{equation}
So finally there are 10 independent transport coefficients 
at second order for some uncharged fluid.

As we have explained above, unless the equations \eqref{relationsintro} are 
satisfied, the fluid dynamics equations do not have a positive divergence 
entropy current. When the equations \eqref{relationsintro} are satisfied
the fluid equations are compatible with the existence of a positive divergence
entropy current, but this current is not unique. Consider a fluid with 
a particular constitutive relation that obeys the equations 
\eqref{relationsintro}. It turns out that any such fluid has a 7 parameter (7 arbitrary functions 
of temperature)
set of positive divergence entropy currents. 

 \begin{equation}\label{duientintro}
 \begin{split}
& J^\mu|_{\text{upto 2nd order}}\\
 =&~su^\mu + \nabla_\nu\left[A_1(u^\mu\nabla^\nu T - u^\nu \nabla^\mu T)\right] + \nabla_\nu \left( A_2 T \omega^{\mu\nu}\right)\\
 & + A_3 \left(R^{\mu\nu} - \frac{1}{2}g^{\mu\nu} R\right) u_\nu
+\left(\frac{A_3}{T} + \frac{dA_3}{dT}\right)\left[\Theta \nabla^\mu T - P^{ab}(\nabla_b u^\mu)( \nabla_a T )\right]\\ &+(B_1\omega^2 + B_2\Theta^2 + B_3 \sigma^2)u^\mu + B_4\left[(\nabla s)^2
u^\mu + 2 s \Theta \nabla^\mu s\right]\\
 \end{split}
 \end{equation}

Our results above apply to an arbitrary theory, i.e. a theory with an 
arbitrary equation of state. The specialization of our results 
to the special case of conformal fluids turns out to be trivial, 
as we now explain. As was explained in (\cite{Baier:2007ix,Loganayagam:2008is}), 
the requirement
of Weyl invariance forces 10 linear combinations of the 15 symmetry allowed 
coefficients 
 to vanish; specifically
\begin{equation}\label{zeros}
\begin{split}
& \kappa_1 = 2\kappa_2\equiv \kappa, ~~\tau = 3\lambda_0,~~\lambda_4 =0\\
&\zeta_1= \zeta_2 = \xi_1 = \xi_2= \xi_3 =0
\end{split}
\end{equation}
Moreover, the the temperature dependence of the 
remaining five coefficients (those that are allowed to take arbitrary values 
consistent with Weyl invariance) is determined by dimensional analysis; 
All of them are just linearly proportional to temperature.

It turns out that this linear dependence on temperarture, 
the conformal equation of state
 and  \eqref{zeros} reduce all of the equations
\eqref{relationsintro} to trivial identities of the form $0=0$. In 
other words, the requirement of positivity invariance of the entropy 
current does not impose any equations on the five  transport
 coefficients (allowed by conformal symmetry).

The coefficients for a conformally covariant entropy 
current are given by the following expressions.
\begin{equation}\label{conformalintro}
\begin{split}
A_1(T) = a_1 ,~~&~~A_2(T) = a_2 ,~~~~A_3(T) = \frac{a_1}{2} T\\
B_1(T) =b_1 T, ~~&~~B_2(T)=\frac{2a_1}{9} T,~~~~B_3(T) =b_3 T\\
&~~B_4(T)= -\left(\frac{a_1}{18}\right)T^{-5}
\end{split}
\end{equation}
where all $a_i$ and $b_i$ are constants.
\newline
Therefore the conformally covariant entropy current has 
four independent coefficients ($a_1, ~a_2,~b_1$ 
and $b_3$) when expanded upto second order in derivatives.

Let us end this introduction with a description of our motivations in
undertaking the computations described in this note. Our first motivation
is practical. Theoretical reconstructions of the RHIC and LHC heavy 
ion experiments often model the expansion of the hot dense deconfined 
plasma by the equations of fluid dynamics including second order 
corrections.  Given that 
we have not been able, from first principles, to compute the fluid
description of QCD it seems of interest to parameterize the most 
general set of allowed equations, as we have done (to 2nd order) in this 
note. 


However our main motivation in undertaking the computations described 
in this note are structural. At zero temperature, the equations of motion 
of physical systems are strongly constrained by the requirement that 
they follow from the extremization of an action. On the other hand 
the tradiational formulation of the equations of fluid dynamics is at the 
level of the equations of motion. It seems likely that the equations 
of fluid dynamics inherit constraints that are the analogue of the 
zero temperature requirement of being obtained from an action. One possible 
source of such constraints, as described in this note, stems from the 
requirment that our system admit an entropy current of positive divergence.
There are other potential sources of constraints, for instance the 
requirement that correlation functions computed from fluid dynamcs 
have certain symmetry properties that can be derived, on general grounds, 
in quantum field theories (see \cite{Jensen:2011xb}). It appears to us to be of interest to find 
a complete `theory' of fluid dynamics; a formalism that ennumerates all 
consistency conditions on the equations of fluid dynamics. Such a formalism
would take the place of the zero temperature requirement that the equations 
of motion follow from an action. The computations presented in this 
note may be thought of as a small first step towards this larger goal.

\section{Brief Summary of our Procedure}\label{sec:summary}

In the rest of this note we proceed to determine the most general 
second order constitutive relations and second order 
entropy current consistent with positivity of the divergence of the current. 
In order to do this we first list out the most general symmetry allowed
entropy current upto third order in the derivative expansion (onshell 
equivalent currents are not treated as distinct). We then perform a 
brute force computation of the divergence of this entropy current, keeping 
all relevant terms (see below for an explaination of which terms are 
relevant) to fourth order in the derivative expansion. We then use 
the equations of motion (including constitutive terms upto second order in 
derivatives) to rewrite our final answer entirely as a function of 
derivatives of velocity and temperature and the background metric 
that are independent of each other  (i.e. are not related 
to each other by the equations of motion). We then work out the 
conditions on the entropy current and constitutive relations that 
ensure the positivity  of this divergence for arbitrary values of the 
independent fluid derivatives, finally obtaining \eqref{relationsintro}
and \eqref{duientintro}

In order to implement the programme outlined in the paragraph above, in 
this note we proceed in the following order. In section \ref{sec:class} below 
we classify and ennumerate the onshell independent derivatives of fluid 
fields (upto fourth order in derivatives). We also enumerate t
all the products of these derivative fields with net derivative number 
$\leq 4$, organizing our ennumeration in representations of the local 
$SO(3)$ that leaves the fluid velocity fixed. 

In section \ref{sec:enropycurrent} below we then proceed to ennumerate the most general 
entropy current upto third order in the derivative expansion. We then 
compute the divergence of this entropy current and determine several 
constraints on the entropy current that follow from the requirement that 
its divergence is positive definite.

In section \ref{sec:finalconstraint} below we then ennumerate the constraints on 
constitutive 
relations, upto second order, that follow from the requirement of positivity 
of divergence of the entropy current. 

We end this brief section by listing our conventions. Throughout this 
note we work in the Landau gauge. In this gauge the velocity $u^\mu$ at any 
point is defined as the unique time-like eigenvector of the stress tensor, 
normalised so that $u_\mu u^\mu = -1$. In other words, by definition
\begin{equation}\label{udef}
T_\mu^\nu u_\nu =-\epsilon u_\mu
\end{equation} 
The quantity $\epsilon$ is taken by definition to be energy density of our  
fluid. All other fluid thermodynamical quantities (like the temperature or 
pressure) are obtained from $\epsilon$ using thermodynamics and equation of state. 
Equation of state expresses the energy density $\epsilon$ as 
a function of some thermodynamic parameter like entropy density and it 
can vary from system to system . In this note we shall keep it arbitrary.
Once $\epsilon(s)$ is known, the temperature $(T)$ and the pressure $(P)$ can be determined 
in the following 
way.
$$T(s) = \frac{d\epsilon(s)}{ds},~~~P(s) = s\frac{d\epsilon(s)}{ds} - \epsilon(s)$$
Both of the above relations directly follow from equilibrium thermodynamics.

\section{Classification of fluid data}\footnote{This section has been 
worked out in collaboration with Shiraz Minwalla and Tarun Sharma.}\label{sec:class}
In this section we present a partial listing of the onshell independent 
derivatives of fluid fields ($T$ and $u^\mu$), at any given point $x$, 
upto fourth order in the derivative expansion. We organize these 
derivatives (which we will often refer to below as independent data) 
according their transformation properties under the $SO(3)$ rotational 
group that leaves $u^\mu(x)$ invariant. 

In order to explain what we mean let us consider a listing of independent 
data at first order in the derivative expansion. Before accounting for 
onshell equivalences we have 16 independent pieces of first derivative data; 
(the four derivatives of temperature and the four derivatives of each of the 
three independent velocities). These 16 pieces of data transform, under 
the local $SO(3)$, as two scalars, two vectors, a pseudo vector and 
a traceless symmetric tensor (i.e. the {\bf 5}) of $SO(3)$ 
(see the second column of Table \ref{table:1storder} for details). However these 16
pieces of data are not all independent. The four {\it perfect} fluid 
equations of motion may be used to solve for four of these fluid derivatives 
in terms of the other 12. As the four fluid equations can be decomposed into 
a vector and a scalar of $SO(3)$ (see the third column of Table 
\ref{table:1storder}) 
it follows that the independent data consists of one vector, one scalar, 
a pseudo vector and a traceless symmetric tensor (see the fourth column of
Table \ref{table:1storder}). The choice of the independent scalar and vector piece 
of data
is arbitrary; we could take our independent data to be either of the vectors 
and either of the scalars listed in the second column of Table 
\ref{table:1storder}. In the 
fourth column of Table \ref{table:1storder} we have made one particular 
choice of the 
independent data that we will employ in much of our note. Occasionally 
we will find it more convenient to use another choice of independent data; we will 
explicitly point this out when this is the case. 

In this note we will require that the production of entropy is positive for 
an arbitrary fluid flow on an arbitrary curved 
manifold. As explained in \cite{Bhattacharya:2011tra}  this requirement yields constraints 
for the constitutive relations of fluids even in the flat space. In order 
to implement the constraint described above, we will find it necessary to 
list the data assocaited with local background metric curvatures in addition 
to the data associated with fluid flows. All curvature invariants formed 
from a background metric are given in terms of (contractions of) the Riemann 
tensor and its derivatives. It is important to recall that, in addition to 
certain symmetry properties, the Reimann tensor also obeys a Bianchi type 
identity. The independent derivatives of the Reimann tensor should be counted
modulo the Bianchi identity and its derivatives. In analogy with the counting
problem for fluid data listed above, we will regard the set of all derivatives
of the Reimann tensor (with all symmetries imposed) as raw data, and 
the Bianchi identities and its derivatives as `equations of motion'. 
We will then list the indpendent pieces of data in curvature and derivatives 
by subtracting equations of motion from raw data, just as described 
in the previous paragraph. 

\subsection{Independent Data}

With no further ado we simply proceed to list the (relevant parts of) 
 fluid and curvature data at various orders in the 
derivative expansion. 

At first order in derivatives we have only fluid data. They are listed below in Table \ref{table:1storder}.
\begin{table}[ht]
\caption{Data at 1st order in derivative} 
\vspace{0.5cm}
\centering 
\begin{tabular}{|c| c| c| c|} 
\hline\hline 
 &Before imposing eom & Eoms & Independent data \\ [1ex] 
\hline 
\hline
Scalars (1) & $ (u.\nabla)T$,~ $(\nabla.u)$ & $u_\nu\nabla_\mu T^{\mu\nu}=0$ & $\Theta \equiv(\nabla.u)$  \\ [0.5ex]  
\hline
Vectors (1) &$(u.\nabla)u^\mu$,~$P^{\mu\nu}\nabla_\nu T$ & $P^\mu_a\nabla_\nu T^{\nu a}=0$ 
& ${\mathfrak a}^\mu \equiv(u.\nabla)u^\mu$ \\[0.5ex]
\hline
Pseudo-vectors (1) & $u_\nu\epsilon^{\nu\mu\lambda\sigma}\nabla_\lambda u_\sigma$ &
 &$ l^\mu \equiv u_\nu\epsilon^{\nu\mu\lambda\sigma}\nabla_\lambda u_\sigma$ \\[0.5ex]
\hline
Tensors (1)& $\nabla_{\langle\mu}u_{\nu\rangle}$
 &  & $\sigma_{\mu\nu}\equiv  \nabla_{\langle\mu}u_{\nu\rangle}$ \\[0.5ex]
\hline
\hline
\end{tabular}
\label{table:1storder} 
\end{table}

Here for any tensor $A_{\mu\nu}$, the symbol $A_{\langle\mu\nu\rangle}$ means the symmetric
 traceless part of 
it, projected in the direction perpendicular to $u^\mu$.
$$A_{\langle\mu\nu\rangle}\equiv   P_\mu^ a P_\nu ^b\left[\left(\frac{A_{ab} +A_{ba}}{2}\right)
 - g_{ab}\left(\frac{P^{\alpha\beta}A_{\alpha\beta}}{3}\right)\right]$$
For example, if we expand this notation, the shear tensor 
$\sigma_{\mu\nu}$ has the following definition
$$\sigma_{\mu\nu}\equiv  P_\mu^ a P_\nu ^b\left[\frac{\nabla_a u_b + \nabla_b u_a}{2}
 - g_{ab}\frac{\Theta}{3}\right]$$

At second order we have curvature data along with fluid data. The curvature 
data is given by the 20 independent components of the Reimannn curvature 
subject to the identities
\begin{equation*}
\begin{split}
&R_{\mu\nu\alpha\beta} = -R_{\nu\mu\alpha\beta} = - R_{\mu\nu\beta\alpha}\\
&R_{\mu\nu\alpha\beta} = R_{\alpha\beta\mu\nu}\\
&R_{\mu[\nu\alpha\beta]}=0
\end{split}
\end{equation*}
The 20 independent components may be decomposed into $SO(3)$ representations as
in Table \ref{table:2ndcurv}
\\

\begin{table}[ht]
\caption{$I_2$ type curvature data} 
\vspace{0.5cm}
\centering 
\begin{tabular}{|c| c| } 
\hline
 &  $R\equiv {R^{\mu\nu}}_{\mu\nu},$\\
Scalars (2)&$R_{00}\equiv u^\mu u^\nu R_{\mu\nu}$\\
&$~~~~~~~\equiv u^\mu u^\nu {R^\alpha}_{\mu\alpha\nu}$\\ [0.5ex]  
\hline
Vectors(1) &$ P^{\mu a}R_{ab} u^b$\\[0.5ex]
\hline
& $R_{\langle \mu\nu\rangle}$\\
Tensors(2) &$F_{\langle\mu\nu\rangle}$\\
& where
 $F_{\mu\nu} \equiv u^\alpha u^\beta R_{\mu\alpha\nu\beta}$\\[0.5ex]
\hline
Pseudo-tensor&$u_b{ R_{\langle \mu }}^{bcd} ~\epsilon_{\nu\rangle c d q}u^q$\\
\hline
\hline
\end{tabular}
\label{table:2ndcurv} 
\end{table}

Second order fluid data is tabulated in Table \ref{table:2ndorder}
\begin{table}[ht]
\caption{$I_2$ type fluid data} 
\vspace{0.5cm}
\centering 
\begin{tabular}{|c| c| c| c|} 
\hline\hline 
 &Before imposing eom & Eoms &Independent data \\ [1ex] 
\hline 
\hline
Scalars (1) &  $(u.\nabla)\Theta,~
\nabla^2 T$& $u_\nu (u.\nabla)\nabla_\mu T^{\mu\nu}=0,$ &
 $(u.\nabla)\Theta$  \\
&$~u^\mu u^\nu \nabla_\mu\nabla_\nu T$&$\nabla_\mu\nabla_\nu T^{\mu\nu}=0$&\\ [0.5ex]  
\hline
Vectors (2) &$P^{\mu\nu}(u.\nabla){\mathfrak a}_\nu,~P^{\mu\nu}\nabla^2 u_\nu,$
& $P^\mu_a(u.\nabla)\nabla_\nu T^{\nu a}=0,$ 
&  $P^{\mu a}\nabla_a \Theta,$ \\
&$P^{\mu\nu}\nabla_\nu \Theta,~P^{\mu\nu}\nabla_\nu (u.\nabla)T$
&$u_aP^{\mu b}\nabla_b\nabla_\nu T^{\nu a}=0$&$P^\mu_a\nabla_b\sigma^{ab}$\\[0.5ex]
\hline
Pseudo-vectors (0)& $(u.\nabla)l^\mu$
 &$u_\nu\epsilon^{\mu\nu\lambda\sigma}\nabla_\lambda\nabla_aT^a_\sigma =0$
 & \\[0.5ex]
\hline
Tensors (1)& $P^{\mu a}P^{\nu b}(u.\nabla)\sigma_{ab},
~\nabla_{\langle\mu}\nabla_{\nu\rangle}T$
 &$\nabla_{\langle\mu}\nabla_aT^a_{\nu\rangle}=0$ 
 & $P^{\mu a}P^{\nu b}(u.\nabla)\sigma_{ab}$ \\[0.5ex]
\hline
Pseudo-tensors (1)&$\nabla_{\langle\mu}l_{\nu\rangle}$&&$\nabla_{\langle\mu}l_{\nu\rangle}$\\[0.5ex]
\hline
Spin-3 (1)&$\nabla_{\langle\mu}\nabla_\nu u_{\alpha\rangle}$&
&$\nabla_{\langle\mu}\nabla_\nu u_{\alpha\rangle}$\\[0.5ex]
\hline
\hline
\end{tabular}
\label{table:2ndorder} 
\end{table}

Third order fluid data that transforms in the scalar, vector and pseudo vector
representations is tabulated in Table \ref{table:3rdfluid} (we will never 
need 3rd order data in the ${\bf 5}$, ${\bf 7}$ and ${\bf 9}$ 
representations, and so do not bother to tabulate these below).
\begin{table}[ht]
\caption{$I_3$ type fluid data} 
\vspace{0.5cm}
\centering 
\begin{tabular}{|c| c| c| c|} 
\hline\hline 
 &Before imposing eom & Eoms & Independent data \\ [1ex] 
\hline 
\hline
Scalars (1) &  $(u.\nabla)^2\Theta,~\nabla^2\Theta,$&
$u_\nu (u.\nabla)^2 \nabla_\mu T^{\mu\nu}=0$,  
 & $(u.\nabla)^2\Theta$  \\
& $(u.\nabla)^3 T,~(u.\nabla)\nabla^2 T$&$(u.\nabla)\nabla_\mu\nabla_\nu T^{\mu\nu}=0$,&
 \\ 
&&$u_\nu \nabla^2\nabla_\mu T^{\mu\nu}=0$&\\[0.5ex]  
\hline
Vectors (1) & $ P^{\mu a} (u.\nabla)^3 u_a$ 
 & $P^{\mu a}u^b\nabla_a\nabla_\mu T^\mu_b =0$
& $P^{\mu a}(u.\nabla) \nabla_a\Theta$ \\
& $P^{\mu a}(u.\nabla) \nabla_a\Theta$
& $P^\mu_a \nabla^2\nabla_\nu T^{\nu a}=0$
 &\\
& $ P^{\mu a}\nabla^2\nabla_a T$&$P^{\mu a}\nabla_a \nabla_\alpha\nabla_\beta T^{\alpha\beta}=0$
&\\
&$ P^{\mu a}(u.\nabla)^2\nabla_a T$& $P^\mu_a (u.\nabla)^2\nabla_\nu T^{\nu a}=0$&\\
&$P^{\mu a} (u.\nabla)\nabla^2 u_a$&&\\[0.5ex]
\hline
Pseudo-vectors (1) & $P^{\mu a}(u.\nabla)^2 l_a$  &
$u_\nu \epsilon^{\mu\nu\alpha\beta} (u.\nabla)\nabla_\alpha\nabla_a T^a_\beta=0$
 &$P^{\mu a}\nabla^2 l_a$ \\
&$P^{\mu a}\nabla^2 l_a$&&\\[0.5ex]
\hline
\hline
\end{tabular}
\label{table:3rdfluid} 
\end{table}

The third order curvature data consists of derivatives of the Reimann curvature
constrained by Bianchi identity 
$$\epsilon^{abcd}\nabla_b R_{\alpha\beta c d}=0$$
In Table \ref{table:3rdcurv} we list the independent curvature 
data that transforms in the scalar, 
vector and pseudo vector representations (again we will not need and so do 
not list the remaining representations)
\begin{table}[ht]
\caption{$I_3$ type curvature data} 
\vspace{0.5cm}
\centering 
\begin{tabular}{|c| c| c| c|} 
\hline\hline 
 &Before imposing eom & Eoms & Independent data \\ [1ex] 
\hline 
\hline
Scalars (2) & $(u.\nabla)R$
&
$u^\mu\epsilon_{\mu a\alpha\beta}\epsilon^{abcd}\nabla_b{ R^{\alpha\beta}}_{ c d}=0$
 &  $(u.\nabla)R_{00}$
\\
&  $ (u.\nabla)R_{00}$ && $(u.\nabla) R$ 
 \\ 
&$u_a\nabla_\mu R^{a\mu}$&&\\[0.5ex]  
\hline
Vectors (3) & $ P^{\mu a}\nabla_a R_{00}$ 
 &$u_a u^\nu\epsilon_{\mu \nu\alpha\beta}\epsilon^{abcd}\nabla_b{ R^{\alpha\beta}}_{ c d}=0 $
&$ P^{\mu a}\nabla_a R_{00}$
 \\
& $P^{\mu a}\nabla_a R$
& $u_\alpha u^\nu\epsilon_{\mu \nu a\beta}\epsilon^{abcd}\nabla_b{ R^{\alpha\beta}}_{ c d}=0 $
 &$P^\mu _a \nabla_\nu R^{\nu a}$\\
& $P^\mu _a\nabla_\nu F^{\nu a}$&&$P^{\mu a}u^b (u.\nabla)R_{ab}$\\
&$P^\mu _a \nabla_\nu R^{\nu a}$& &\\
&$P^{\mu a}u^b (u.\nabla)R_{ab}$&&\\[0.5ex]
\hline
Pseudo-vectors (1) & $u^p u_a \epsilon^{abcd}\nabla_b {R^\mu }_{p cd}$  &
$u_\alpha u_a\epsilon^{abcd}\nabla_b{ R^{\alpha\mu}}_{ c d}=0$
 &$u^p u_a \epsilon^{\mu abc}\nabla_b R_{p c}$ \\
&$u^p u_a \epsilon^{\mu abc}\nabla_b R_{p c}$&&\\[0.5ex]
\hline
\hline
\end{tabular}
\label{table:3rdcurv} 
\end{table}

Finally, fourth order scalar data (all we will need at fourth order), both 
fluid as well as curvature, is tabuated in Table \ref{table:4th}.

\begin{table}[ht]
\caption{$I_4$ type scalars} 
\vspace{0.5cm}
\centering 
\begin{tabular}{|c| c| c| c|} 
\hline\hline 
 &Before imposing eom & Eoms & Independent data \\ [1ex] 
\hline 
\hline
Fluid data (1) & $(u.\nabla)^3\Theta$
&
$(u.\nabla)^3\left(u_\nu\nabla_\mu T^{\mu\nu}\right)=0$
 &  $(u.\nabla)^3\Theta$
\\
&  $ (u.\nabla)\nabla^2\Theta$ &$(u.\nabla)\nabla^2\left(u_\nu\nabla_\mu T^{\mu\nu}\right)=0$
&
 \\ 
&$(u.\nabla)^4 T$&$(u.\nabla)^2\nabla_\mu\nabla_\nu T^{\mu\nu}=0$&
\\
&$(u.\nabla)^2\nabla^2 T$&$\nabla^2\nabla_\mu\nabla_\nu T^{\mu\nu}=0$&
\\
&$\nabla^2 (\nabla^2 T)$&&\\[0.5ex]  
\hline
Curvature data (4) & $(u.\nabla)^2R_{00}$ 
 &$u_a u^\nu\epsilon_{\mu \nu\alpha\beta}\nabla^\mu\epsilon^{abcd}\nabla_b{ R^{\alpha\beta}}_{ c d}
=0 $
&$(u.\nabla)^2R_{00}$
 \\
& $(u.\nabla)^2 R$
& $u_\alpha u^\nu\epsilon_{\mu \nu a\beta}\nabla^\mu\epsilon^{abcd}\nabla_b{ R^{\alpha\beta}}_{ c d}
=0 $
 &$(u.\nabla)^2R_{00}$
\\
& $\nabla^2 R$&$u^\mu\epsilon_{\mu a\alpha\beta}(u.\nabla)
\epsilon^{abcd}\nabla_b{ R^{\alpha\beta}}_{ c d}=0$
&$\nabla^2 R$
\\
&$\nabla^2 R_{00}$& &$\nabla^2 R_{00}$
\\
&$u_a (u.\nabla) \nabla_b R^{ab}$&&\\
&$\nabla_a\nabla_b R^{ab}$&&\\
&$\nabla_a\nabla_b F^{ab}$&&\\[0.5ex]
\hline
\hline
\end{tabular}
\label{table:4th} 
\end{table}

\subsection{Composite Expressions}

In the sequel we will sometimes need to list for example, 
all  3rd order vectors. In addition to expressions constructed
out of the independent 3rd order data, listed in the previous subsection, 
the set of all 3rd order vectors includes expressions cubic in first order 
data, and expressions formed out of the product of one first order and one 
second order piece of data. We will refer to expressions constituted out 
of products of independent data as composite expressions. 
Composite expressions formed out of independent 
data are easily ennumerated and decomposed into $SO(3)$ representations 
using Clebsh Gordan decompositions (taking care to acccount for 
symmetry properties when we multiply two or three copies of the same data). 

In order to ease the process of reference to composite expressions in the 
rest of the note we now adopt the following terminology. Independent 
data at $m^{th}$ order in the derivative expansion is referred to data of 
the type $I_{m}$. A composite expression that consists of a product of 
three first order pieces of data is referred to as an expression of the type 
$C_{1,1,1}$. Composite expressions that consist of the product of a first order
and 3rd order piece of data are called expressions of the form $C_{1,3}$. 
The generalization of our notation to other forms of composite data is 
obvious. 

In the sequel we will need to list only those composite expressions that 
transform in the vector (in order to list the most general entropy current) 
or the scalar (in order to list the most general terms in its divergence). 
In the rest of this subsection we present a listing of those vector and scalar 
composite expressions that will be needed below.

\begin{table}[ht]
\caption{$C_{1,1}$ type expressions} 
\vspace{0.5cm}
\centering 
\begin{tabular}{|c| c|} 
\hline\hline 

\hline
Scalars (4) &$\Theta^2,~~{\mathfrak a}^2,~~\omega^2,~~\sigma^2$
\\[0.5ex]
\hline
Vectors (3) &${\mathfrak a}^\mu \Theta,~~{\mathfrak a}_\nu \omega^{\mu\nu},
~~{\mathfrak a}_\nu \sigma^{\mu\nu}$\\[0.5ex]
\hline
Tensors (5) &$\Theta \sigma_{\mu\nu}$,
~ $\sigma_{\langle\mu}^a\sigma_{a\nu\rangle}$,~$\omega_{\langle\mu}^a\sigma_{a\nu\rangle}$,~
$\omega_{\langle\mu}^a\omega_{a\nu\rangle}$,
~${\mathfrak a}_{\langle \mu}{\mathfrak a}_{\nu\rangle}$\\[0.5ex]
\hline
\hline
\end{tabular}
\label{table:c11} 
\end{table}
Here $\omega_{\mu\nu} =  P^{\mu a} P^{\nu b}\left[\frac{\nabla_a u_b - \nabla_b u_a}{2}  \right]$

\begin{table}[ht]
\caption{$C_{1,2}$ type expressions independent of the curvature} 
\vspace{0.5cm}
\centering 
\begin{tabular}{|c| c|} 
\hline\hline 

\hline
Scalars (4) &$\Theta (u.\nabla)\Theta,~~(\nabla_\mu T)\nabla^2 u^\mu,
~~(\nabla_\mu T)(u.\nabla)(\nabla^\mu T),~~\sigma_{\mu\nu}\nabla^\mu\nabla^\nu T$
\\[0.5ex]
\hline
\hline
Vectors (11) &$P^\alpha_\mu\Theta \nabla^2 u^\mu,~~P^\alpha_\mu\Theta (u.\nabla)\nabla^\mu T,
~~P^\alpha_\mu(\nabla^2 T)\nabla^\mu T,~~\omega_{\mu\nu}\nabla^2 u^\mu$\\
&$\omega_{\mu\nu} (u.\nabla)\nabla^\nu T,~~\sigma_{\mu\nu}\nabla^2 u^\mu,~~
\sigma_{\mu\nu} (u.\nabla)\nabla^\nu T,~~P^\alpha_\mu(\nabla_a T)(\nabla^\mu\nabla^a T)$\\
&$(\nabla _a \omega_{b\mu})\sigma^{ab},~~ P^\alpha_\mu\sigma_{ab}\nabla^a\nabla^b u^\mu,~~
P^\alpha_\mu\omega ^{ab}\nabla^\mu \omega_{ab}$\\[0.5ex]
\hline
\hline
\end{tabular}
\label{table:c12} 
\end{table}

\begin{table}[ht]
\caption{$C_{1,2}$ type expressions involving a curvature} 
\vspace{0.5cm}
\centering 
\begin{tabular}{|c| c|} 
\hline\hline 

\hline
Scalars (5) &$F_{ab}\sigma^{ab},~~R_{ab}\sigma^{ab},
~~u_a{\mathfrak a}_bR^{ab},~~\Theta R,~~\Theta R_{00}$
\\[0.5ex]
\hline
Vectors (9) &$P^\alpha_\mu{\mathfrak a}_\nu F^{\mu\nu},~~P^\alpha_\mu{\mathfrak a}_\nu R^{\mu\nu},
~~{\mathfrak a}^\mu R_{00},~{\mathfrak a}^\mu R,~~P^\alpha_\mu u_a\Theta R^{a\mu}$\\
&$u^a R_{ab}\sigma^{b\mu},~~u_a R^{a\mu bc}\omega_{bc},~~
P^\alpha_\mu u_a R^{ab \mu c}\sigma_{bc},~~u^a R_{ab}\omega^{b\mu}$\\[0.5ex]
\hline
\hline
\end{tabular}
\label{table:c12curv} 
\end{table}

\begin{table}[ht]
\caption{$C_{1,1,1}$ type expressions} 
\vspace{0.5cm}
\centering 
\begin{tabular}{|c| c|} 
\hline\hline 

\hline
Scalars (7) &$\Theta^3,~~\sigma^2\Theta,
~~\omega^2 \Theta,~~{\mathfrak a}^2\Theta$
\\
& ${\mathfrak a}_\mu {\mathfrak a }_\mu \sigma^{\mu\nu},~~\sigma_{\mu a}\sigma^a_b\sigma^{b\mu},
~~\omega_{\mu a}\sigma^a_b\omega^{b\mu}$\\[0.5ex]
\hline
Vectors (10) &$\sigma^2 {\mathfrak a}^\mu,~~\omega^2 {\mathfrak a}^\mu,
~~\Theta^2 {\mathfrak a}^\mu,~~{\mathfrak a}^2 {\mathfrak a}^\mu,
~~\Theta \sigma^{\mu\nu}{\mathfrak a}_\nu$\\
&$\Theta \omega^{\mu\nu}{\mathfrak a}_\nu,~~\sigma^{\mu a}\sigma_{ab}{\mathfrak a}^b,~~
~\omega^{\mu a}\omega_{ab}{\mathfrak a}^b,~~\omega^{\mu a}\sigma_{ab}{\mathfrak a}^b
,~~\sigma^{\mu a}\omega_{ab}{\mathfrak a}^b$\\[0.5ex]
\hline
\hline
\end{tabular}
\label{table:c111} 
\end{table}

\section{Entropy current}\label{sec:enropycurrent}

In this section we will derive constraints on constitutive relations, at 
second order in the derivative expansion, from the requirement of positivity of 
divergence of {\it any} entropy current that reduces to $s u^\mu$ (where 
$s$ is the entropy density) in equilibrium. 

We will first explain in very broad terms how we proceed. 

The entropy current takes the form 
\begin{equation}\label{mpec}
J^\mu = J^\mu_{eq} + \tilde J^\mu
\end{equation}
where 
$$J^\mu_{eq} = s u^\mu$$
In general ${\tilde J}^\mu$ has terms of all orders in the derivative 
expansion, but for the purposes of this note we will find it sufficient 
to truncate ${\tilde J}^\mu$ to terms of third order or lower in 
derivatives. To start with we allow ${\tilde J}$ to be given by the most 
general possible form consistent with symmetries. We then compute the divergence 
of $J^\mu$ and reexpress 
the final result entirely in terms of independent data of fourth or
lower order in derivatives. The last step (reexpressing the divergence 
of $J$ in terms of independent data) uses the equations of fluid dynamics, 
and so the constitutive relations.  Our final expression is a polynomial 
in the (finite number of) pieces of data of fourth or lower order in 
derivatives. We then demand that the resultant polynomial is positive 
definite (or can be made so by the addition of terms higher than fourth
order in the derivative expansion) as a function of its arguments. 
This rather stringent requirement turns out to yield several constraints 
on the form of both the entropy current at second (and third) order 
as well as constitutive relations at second order in the derivative expansion. 

\subsection{Entropy current in equilibrium and 1st order correction}
In order to set the stage for our discussion we first recall how the 
requirement of positivity of the entropy current constrains constitutive
relations at first order in the derivative expansion \cite{Bhattacharya:2011tra}.
We first recall that thermodynamics and the fluid dynamical equations 
may be used to demonstrate that 
\begin{equation}\label{divperf}
\nabla_\mu J^\mu_{eq} = -\frac{1}{T}\left(\Pi^{\mu\nu}\sigma_{\mu\nu} 
+\frac{ \Theta\Pi^\mu_\mu }{3}\right)
\end{equation}
It follows in particular from \eqref{divperf} that entropy is conserved in 
perfect fluid dynamics (i.e. when $\Pi^{\mu\nu}$ vanishes). It also follows 
that the divergence of the most general entropy current \eqref{mpec} 
only contains terms of second or higher order in the derivative expansion. 
Let us now examine the constraints from the requirement of positivity of 
these second order pieces. For this purpose we need to study the most 
general entropy current at first order in derivatives. Imposing the 
requirement of invariance under partiy, the most general (onshell inequivalent)
family of first order entropy currents is given by 
\begin{equation} \label{gfec}
J^\mu=J^\mu_\text{equilibrium} + \alpha \Theta u^\mu +\beta {\mathfrak a}^\mu
\end{equation}
(we have used here that at first order in the derivative expansion we 
have one piece of scalar data,  which may be chosen as $\Theta$, and one
piece of vector data, which may be chosen as ${\mathfrak a}^\mu$). 
We now proceed to compute the divergence of \eqref{gfec} and
use the perfect fluid equations to rexpress the result in terms of 
independent data. The resultant expression is the sum of a quadratic 
form in first order data and a linear form in 2nd order scalar data. 
As derived in \cite{Bhattacharya:2011tra}, the final expression for this divergence
 is given as
\begin{equation}\label{diveq1}
 \begin{split}
  &\nabla_\mu J^\mu|_{\text{upto 2nd order}}\\
=&   -\frac{1}{T}\left(\Pi^{\mu\nu}\sigma_{\mu\nu} +\frac{ \Pi \Theta}{3}\right) \\
&+ \Theta (u.\nabla)\alpha + ({\mathfrak a}.\nabla)\beta 
+ \left(\alpha + \frac{\beta}{3}\right)\frac{\Theta^2}{3} + \beta\left(\sigma^2 + \omega^2\right)\\
&+ (\alpha + \beta) (u.\nabla)\Theta + \beta R_{00}
 \end{split}
\end{equation}

Here both the first and the second line have terms quadratic in 1st order data. The 
last line contains the terms 
which are linear in second order data. There are three independent 2nd order scalars
($(u.\nabla)\Theta,~R,~R_{00}$) as given in the classification in section \ref{sec:class}. 
Only two of these three scalars appear in \eqref{diveq1}.  Since these two terms are 
linear in fluid variable, 
they can have any sign and to ensure positivity of the divergence their coefficients
 (both $\alpha$ and $\beta$) must be set to zero.
This implies that at 1st order no correction can  be added to the entropy current which 
is consistent with the 
positivity requirement. Then in the RHS of \eqref{diveq1} only the first line will give
 a non-zero contribution.
To evaluate the first line we need the first order corrections to the constitutive 
relation. 
At first order the most general correction to the constitutive relation (stress tensor 
in Landau gauge)
 will involve the single on-shell 
independent 1st order scalar which we have chosen to be $\Theta$ and single on-shell 
independent tensor 
$\sigma_{\mu\nu}$.

$$\Pi_{\mu\nu}|_\text{upto 1st order} = -\eta\sigma_{\mu\nu} -\zeta \Theta P_{\mu\nu}$$
where $\eta$ and $\zeta$ are shear and bulk viscosity respectively.

Therefore finally
\begin{equation}\label{diveq2}
 \begin{split}
  &\nabla_\mu J^\mu|_{\text{upto 2nd order}}
=\frac{1}{T}\left(\eta\sigma^2 +\zeta\Theta^2\right) \\
\end{split}
\end{equation}
Hence to have a positive definite divergence one requires  that 
$$\eta\geq0,~~~\zeta\geq0$$

The main point to note in the above equation \eqref{diveq2} is that it involves only 
two of the four first order 
on-shell independent data as listed in section \ref{sec:class}.  The squares of the 
independent vector 
 ${\mathfrak a}^\mu$ and the pseudo-vector $l^\mu$ do not appear in equation
 \eqref{diveq2}. 
Because of this fact any term in the divergence which is 
of the form  (${\mathfrak a}_\mu \times \text{$I_2$ or $I_3$ type vector}$) or 
($l_\mu \times \text{$I_2$ or $I_3$ type pseudo-vector}$) can never be made
 positive-definite.
  
  \subsection{General constraints on second and the third order corrections}
  In general $\tilde J^\mu$ can be written as
 $$\tilde J^\mu = \left(\sum_i {\mathfrak S}_i\right)u^\mu + \sum_i{\mathfrak V}_i^\mu$$
  where ${\mathfrak S}_i$ is an arbitrary combination of $i$th order on-shell independent
 scalars and ${\mathfrak V}^\mu_i$ is a combination of $i$th order vectors.
  In the previous subsection we have seen that to constrain the first order transport 
coefficients $\eta $ and $\zeta$ we need to determine only the first order correction 
to the entropy current (i.e. only ${\mathfrak S}_1$ and ${\mathfrak V}^\mu_1$ and both of
 them finally turn out to be zero). But to constrain the second order transport 
coefficients we need to go till the third corrections to the entropy current. The 
reason is the following.
 
Suppose the divergence of the most general entropy current  has two terms of the form
$$\nabla_\mu J^\mu_s = A x^2 + B x y= Ax^2\left(1 + \frac{By}{Ax}\right) $$
where $x$ and $y$ are two on-shell independent fluid data and $A$ and $B$ are some
 functions of temperature, which in general will depend on the coefficients appearing 
in the entropy current or transport coefficients.

In this schematic expression of divergence since $x$ and $y$ are two independent fluid 
data, locally the ratio $\frac{By}{Ax}$ can take any negative value, larger or smaller 
than 1 in magnitude and the positivity constraint will depend on whether $y^2$ term  is
 present or not in the final expression of the divergence. In the absence of a $y^2$ 
piece, the coefficient $B$ has to be set to zero and the coefficient $A$ to some 
non-negative number.

 But this argument does not require $x$ and $y$ to be of same order in derivative
 expansion.  Even when $x$ is of first order in derivative and $y$ is of second order, 
the ratio $\frac{By}{Ax}$ can be of order 1 for some particular fluid configuration 
where $x$ is accidentally small enough to be comparable to $y$ at a given point. 
 In such cases to see whether $y^2$ term is present or not, we need to compute the 
divergence till fourth order. This is why we have to compute the divergence till 
fourth order even if we want to constrain just the second order transport coefficients.

 In fact the constraints on transport coefficients will involve situation where 
$x$ and $y$ are necessarily of different orders. For example, $B$ will contain some 
second order transport coefficients only when $x$ is of first order (as we will see 
below that $x$ has to be equal to $\sigma_{\mu\nu}$ or $\Theta$) and $y$ is of second order in 
derivatives. It will turn out that most of the equalities among the coefficients will 
follow from this sort of argument.

Below we schematically list all the constraints we need to impose on the third and
 fourth order pieces of the divergence in order to ensure its positivity.
\begin{itemize}
\item The coefficient of any term (appearing in third or fourth order piece of 
the divergence) which contains more than one factors of $\sigma_{\mu\nu}$ or $\Theta$ 
or at least one factor of $\Theta\sigma_{\mu\nu}$ will not have any constraint from 
positivity as long as $\eta$ and $\zeta$ are non-zero and are of order one. This is 
because whenever such third or fourth order terms are non-zero, the second order piece 
of the divergence is also non-zero and positive-definite and will always dominate these
 terms within derivative expansion. These terms can never make the divergence negative.
 Therefore while calculating the the third and fourth order divergence we shall ignore 
all these terms.

\item One needs to do sixth order analysis to constrain the coefficients of any term 
(appearing in the fourth order divergence of the entropy current) which is of the form 
($\sigma_{\mu\nu} \times \text{Some third order tensor}$) 
or ($\Theta \times \text{Some third order scalar}$). Such 
terms generically will have contributions from third order 
transport coefficients. Since we are interested only upto second 
order transport coefficients we shall ignore all such terms while 
calculating the fourth order divergence.

\item The coefficients of all the terms which contain a single
 $I_2$, $I_3$ or $I_4$ type scalar (at second, third and fourth order respectively) 
have to be set to zero. This is because locally all these terms are linear in fluid
 variables and therefore can have any sign.

\item In the third order and fourth order piece of the divergence, the 
coefficients of all the terms which are of the form 
(${\mathfrak a}_\mu \times \text{$I_2$ or $I_3$ type vector}$) or
 ($l_\mu \times \text{ $I_3$ type pseudo-vector}$) have to be set to zero. 
Since there is no on-shell independent $I_2$ type pseudo-vector, there 
could not be any term of the form ($l_\mu \times \text{ $I_2$ type pseudo-vector}$).

\item At this stage the terms appearing in the third order piece of the 
divergence will be of the following form.
\begin{enumerate}
\item $\sigma_{\mu\nu}\times(\text{$I_2$ or $C_{1,1}$ type tensors})$
\item $\Theta\times(\text{$I_2$ or $C_{1,1}$ type scalars})$
\end{enumerate}
All these terms will involve the second order transport coefficients.

 The relevant terms appearing at the fourth order (where all the terms 
involving $\sigma_{\mu\nu}$ and $\Theta$ are ignored) will be of the 
following form.
 \begin{enumerate}
 \item A quadratic form involving independent $I_2$ type data
 \item A quartic form involving ${\mathfrak a}_\mu$ and $\omega_{\mu\nu}$
\item Terms linear in $I_2$ type data and quadratic in ${\mathfrak a}_\mu$ 
and/or $\omega_{\mu\nu}$
 
 \end{enumerate}
 Therefore when $\eta\neq0$ and $\zeta\neq0$ the relevant part of the
 divergence calculated upto fourth order is schematically given by
 
 \begin{equation}\label{schmdiv}
 \begin{split}
 \text{Divergence}= &\frac{\eta~\sigma^2 + \zeta~\Theta^2}{T}\\ 
 &+ \sigma_{\mu\nu}\times(\text{$I_2$ or $C_{1,1}$ type tensors})
 +\Theta\times(\text{$I_2$ or $C_{1,1}$ type scalars})\\
 &+\text{A quadratic form involving independent $I_2$ type data} \\
 &+\text{Terms linear in $I_2$ type data and quadratic in ${\mathfrak a}_\mu$ 
and/or $\omega_{\mu\nu}$}\\
 & +\text{A quartic form involving ${\mathfrak a}_\mu$ and $\omega_{\mu\nu}$}
 \end{split}
 \end{equation}
where in the second line all the $C_{1,1}$ type tensors involving $\sigma_{\mu\nu}$ 
and all the $C_{1,1}$ type scalars involving $\Theta$ are ignored.

\item Now we can shift $\sigma_{\mu\nu}$ by a combination of $I_2$ or $C_{1,1}$ type 
tensors such that the term linear in $\sigma_{\mu\nu}$ appearing in 
the second line gets absorbed.
This shift will generate fourth
 order terms structurally similar to the terms appearing in third, fourth and fifth line 
of the above equation. One can see that all these newly generated terms together will 
necessarily be negative definite. Similar shift has to be done to absorb the terms linear 
in $\Theta$  to the first line of \eqref{schmdiv}.

\item One can do similar 
shifts in $I_2$ type data to absorb the terms appearing in the fourth line of equation
 \eqref{schmdiv} into terms 
appearing in the third and fifth line with $I_2$ data replaced by the shifted one.
At this stage the schematic expression of the divergence will take the following form.
\begin{equation}\label{schmdivp}
 \begin{split}
 \text{Divergence}= &\frac{\eta~(\text{shifted}~\sigma)^2 + \zeta~(\text{shifted}~ 
\Theta)^2}{T}\\ 
  &+\text{A quadratic form involving shifted $I_2$ type data} \\
  & +\text{A quartic form involving ${\mathfrak a}_\mu$ and $\omega_{\mu\nu}$}
 \end{split}
 \end{equation}
 
 \item The positive definiteness of the divergence finally will imply the 
positivity of the quadratic and the quartic form appearing in the second and 
the third line of \eqref{schmdivp}.
 \end{itemize}
 
 Such condition will generically give some inequalities among the coefficients.
However suppose by explicit computation one finds that for some particular 
negative definite 
term generated by the shift
 there is no term present in the third or fifth line 
of equation \eqref{schmdiv} to compensate. Then this will imply that 
the coefficient of the 
corresponding  linear term (the source for generating this particular
negative-definite term through shift) in the second line or fourth
 line has to be set to zero. 
This will give strict equalities among the coefficients.

 It will turn out that all of the constraints on the 2nd order transport 
coefficients will arise from this sort of argument.
 
In explicit calculation we will see that in the quadratic form involving the $I_2$ 
type data there will not be any term 
proportional to $R_{00}^2$, 
$R^2$, $F_{\mu\nu}F^{\mu\nu}$ and $R_{\mu\nu}R_{ab}P^{\mu a}P^{\nu b}$. 
This will imply that the coefficients of all the terms linear in $R_{00}$,
$R$, $F_{\mu\nu}$ and $R_{\mu\nu}P^{\mu a}P^{\nu b}$ have 
to be zero. It turns out that once we set these linear terms to zero,
the quartic form mentioned in
 the last line of equation \eqref{schmdiv} also vanishes.

The vanishing of these terms at fourth order gives the final constraint 
on the transport coefficients. In the explicit computation we will see that there
are eight
 terms ($\Theta {\mathfrak a}^2$, $\Theta~ l^2$,
$\sigma_{\mu\nu}{\mathfrak a}^\mu{\mathfrak a}^\nu$, $\sigma_{\mu\nu}l^\mu l^\nu$ 
$\sigma_{\mu\nu}R^{\mu\nu}$, $\sigma_{\mu\nu}F^{\mu\nu}$, $R\Theta$ and $R_{00}\Theta$)
in the third order divergence which are linear in the set of fluid and curvature
data mentioned above and also involve eight independent transport coefficients. 
So setting the coefficient of these linear terms to zero, we can
express the eight transport coefficients in terms of
 the coefficients appearing in the second order entropy current. It will turn out only 
three of the entropy current coefficients appear in these expressions. Eliminating these
 three coefficients we get the final five relations among the 15 transport coefficients
as presented in \eqref{relationsintro}.

Once all these relations are imposed on the divergence, one is left with a quadratic 
form involving only $I_2$ type data . To ensure that this quadratic form is 
positive-definite the coefficients appearing in the second and third order entropy 
current as well as the transport coefficients have to satisfy some inequalities. 
But in this case, at least upto this order the entropy current coefficients 
can not be eliminated from the relations. Therefore unlike the first order 
transport coefficients the second order ones do not satisfy any inequalities within 
themselves.
 \subsection{Implementing the general rules at second order}
 At second order we have to determine ${\mathfrak S}_2$ and ${\mathfrak V}_2^\mu$ 
such that the divergence calculated upto third order in derivative expansion is 
non-negative. Here we shall follow the general procedure described in  the previous 
subsection.
  
  We shall express ${\mathfrak S}_2$ and ${\mathfrak V}_2^\mu$ in terms of the 
on-shell independent second order scalars and vectors respectively.
 ${\mathfrak S}_2$ will have 7 coefficients, three multiplying the three 
independent $I_2$ type scalars and rest of four multiplying the four $C_{1,1}$ 
type scalars. ${\mathfrak V}^\mu$ will also have 6 coefficients, three multiplying 
the three $I_2$ type vectors and the rest multiplying the three $C_{1,1}$ type vectors. 
 
 So before imposing any constraint the entropy current at second order contains total 
13 coefficients, each of which is an arbitrary function of temperature.
 
We shall write this most general 13 parameter entropy current in the following form .
 \begin{equation}\label{duient1}
 \begin{split}
 \tilde J^\mu|_{\text{second order}} =& \nabla_\nu\left[A_1(u^\mu\nabla^\nu T - u^\nu \nabla^\mu T)\right] + \nabla_\nu \left( A_2 T \omega^{\mu\nu}\right)\\
 & + A_3 \left(R^{\mu\nu} - \frac{1}{2}g^{\mu\nu} R\right) u_\nu
+\left[ A_4 (u.\nabla)\Theta  + A_5 R + A_6 R_{00}\right] u^\mu\\ &+(B_1\omega^2 + B_2\Theta^2 + B_3 \sigma^2)u^\mu + B_4\left[(\nabla s)^2
u^\mu + 2 s \Theta \nabla^\mu s\right]\\
&+\left[\Theta \nabla^\mu B_5 - P^{ab}(\nabla_b u^\mu)( \nabla_a B_5 )\right]+ B_6 \Theta {\mathfrak a^\mu} + B_7 {\mathfrak a}_\nu \sigma^{\mu\nu}
 \end{split}
 \end{equation}
 Here $s$ is the entropy density and all the coefficients $A_i$ and the $B_i$ 
are the arbitrary functions 
of temperature.

Now we shall argue that \eqref{duient1} is actually the most general 13 parameter
 entropy
 current. 
 By equations of motion one can show that the only $I_2$ type vector appearing in 
the first term 
is $P^{\mu a}\nabla_a \Theta$, the second term contains a linear combination of all 
the three independent $I_2$ type vectors and in the third term
 the only $I_2$ type vector that appears is $P^{\mu a} R_{ab}u^b$. Therefore the 
first three terms
 together take care of the all the three $I_2$ type vectors. 
 Terms multiplying $A_4,~~A_5$ and $A_6$ are the three $I_2$ type scalars.
 
 By equation of motion $B_4$ term is equal to a linear combination of $\Theta^2 u^\mu$, 
${\mathfrak a}^2 u^\mu$ and $\Theta {\mathfrak a^\mu}$. Similarly $B_5$ term is a
 particular 
linear combination of $\Theta^2 u^\mu$, ${\mathfrak a}^2 u^\mu$, $\Theta {\mathfrak a^\mu}$,
 ${\mathfrak a}_\nu \sigma^{\mu\nu}$ and ${\mathfrak a}_\nu \omega^{\mu\nu}$. 
Therefore all 
the $C_{1,1}$ type scalars and vectors appear in \eqref{duient1} with distinct 
coefficients.
 
Next we shall compute the divergence of this 13 parameter entropy current constructed in
 \eqref{duient1}. We have to set the coefficients of all the  $I_3$ type  on-shell 
independent terms to zero.
Since there are total 3 independent $I_3$ type scalars, it can impose at most three 
relations among the coefficients 
appearing in the second order entropy current. 
 Next we have to isolate all the $C_{1,2}$ type terms which are of the form of 
${\mathfrak a}_\mu$ times a $I_2$ type 
vector and set their coefficients to zero. 
 Since there are total three second order $I_2$ type vectors this condition also can 
impose at most three constraints.


\begin{itemize}
\item The divergence of the first two terms in \eqref{duient1} vanish identically.
\item The divergence of the third term (the term with coefficient $A_3$) does not 
produce any $I_3$ type scalar. 
The divergence of this term is explicitly computed in \eqref{a3term}.
\item The three independent $I_3$ type scalars 
($u^a u^b \nabla_a \nabla_b \Theta,~~u.\nabla R,~~u.\nabla R_{00}$) are produced
from the three terms multiplying coefficients $A_4$, $A_5$ and $A_6$ respectively. 
Therefore to maintain 
positivity $A_4$, $A_5$ and $A_6$ have to be set to zero.
\item The divergence of the terms multiplying $B_1,~~B_2,~~B_3$ and $B_4$ do not

 produce 
any term of the form ${\mathfrak a}^\mu$ times an $I_2$ type vector.
The divergence of these terms are explicitly computed in \eqref{termb1}, \eqref{b2b3term},
 and \eqref{b4term} respectively.

\item The divergence of the terms with coefficients $B_6$ and $B_7$ produce the 
two terms
 ${\mathfrak a}_\nu\nabla_\mu \sigma^{\mu\nu}$ and ${\mathfrak a}_\mu\nabla^\mu\Theta$ 
respectively whose net coefficient should be zero to ensure the positivity of the 
divergence.

Since these are the only places where these terms are produced, $B_6$ and $B_7$ are
 set 
to zero.
\item Both the terms multiplying $B_5$  and $A_3$ produce the third possible term of
 the 
form ${\mathfrak a}^\mu$ times an $I_2$ type vector which is 
${\mathfrak a}_\mu R^{\mu\nu}u_\nu$ 
(see \eqref{a3term} and \eqref{b5term}). The net coefficient is 
$\left(\frac{A_3}{T} + \frac{dA_3}{dT}-\frac{d B_5}{dT}\right)$. 

Therefore positivity implies
$$\frac{d B_5}{dT}=\frac{A_3}{T} + \frac{dA_3}{dT}$$
\end{itemize}

 After imposing all these constraints the final form of the second order 
entropy current is given as
\begin{equation}\label{duient2}
\begin{split}
&\tilde J^\mu|_{\text{second order}}\\
 =& \nabla_\nu\left[A_1(u^\mu\nabla^\nu T - u^\nu \nabla^\mu T)\right] +
 \nabla_\nu \left( A_2 T \omega^{\mu\nu}\right)\\
 & + A_3 \left(R^{\mu\nu} - \frac{1}{2}g^{\mu\nu} R\right) u_\nu
+\left(\frac{A_3}{T} + \frac{dA_3}{dT}\right)\left[\Theta \nabla^\mu T - 
P^{ab}(\nabla_b u^\mu)( \nabla_a T )\right]\\ &+(B_1\omega^2 + 
B_2\Theta^2 + B_3 \sigma^2)u^\mu + B_4\left[(\nabla s)^2
u^\mu + 2 s \Theta \nabla^\mu s\right]\\
\end{split}
\end{equation}

\subsection{Implementing the general rules at third order}
\begin{itemize}
\item First we have to write ${\mathfrak S}_3$ and ${\mathfrak V}_3^\mu$  
in terms of the on-shell independent data. 
\item The coefficients of all the $I_4$ type data appearing 
at the fourth order divergence have to be set to
 zero. An $I_4$ type term in the fourth order divergence can occur only when 
the derivative acts on the $I_3$ type terms of the third order entropy current.
Therefore this condition will constrain the coefficients of the $I_3$ type terms. 

Now there are 3 $I_3$ type independent third order scalars and 4 $I_3$ type independent
 third order vectors and there are total 5 $I_4$ type fourth order scalars. This means 
one 
can have at least 2 distinct coefficients multiplying $I_3$ type terms in a third order
 entropy current with positive definite divergence.

It turns out that after imposing this constraint there are exactly two coefficients
 left and 
the terms multiplying them can be chosen in such a way that their divergence vanish 
identically. 
So these two terms will not contribute to any further constraint.

\item The number of free coefficients that can multiply the $C_{1,2}$ type data in 
third order entropy
 current is  quite large. There can be total 9 free coefficients in ${\mathfrak S}_3$
 and 20  
in ${\mathfrak V}_3^\mu$.

These coefficients will be constrained by the fact that any terms of the form 
${\mathfrak a}_\mu$
times a $I_3$ type vector or $l_\mu$ times a $I_3$ type pseudo-vector have to be 
set to zero.

Since there are total 4 $I_3$ type vectors and total 2 $I_3$ pseudo-vectors, it 
can produce at most 
 6 constraints, reducing the number of free coefficients to 23.

\item Since there is no other general constraint to simplify the form of the entropy 
current at 
this stage we have to calculate the divergence.

But we will not attempt to calculate the full divergence. Instead we shall calculate 
only those 
terms which can impact the constraints on the second order transport coefficients.

For this purpose in the divergence we can ignore all those terms which are multiplied 
by $\Theta$ 
or $\sigma_{\mu\nu}$.

Also to simplify we shall try to write the terms in the form of ($\nabla_\mu A^{\mu\nu}$ ) 
where $A^{\mu\nu}$ is an anti-symmetric tensor, so that their divergence vanish 
identically. 
We could not do it for all the independent terms, but we try to apply this trick for 
as many terms as possible.
\end{itemize}

Now we shall explicitly construct the required part of the third order entropy current
 piece by piece.

The part multiplying the $I_3$ type terms can be written as
\begin{equation}\label{tinent1}
\begin{split}
&\tilde J^\mu|_{\text{3rd order/$I_3 $type}}\\
 =& \nabla_\nu \left[P_1(u^\nu (u.\nabla) \nabla^\mu T - 
u^\mu (u.\nabla)\nabla^\nu T)\right] + 
\nabla_\nu \left[P_2(u^\nu R^\mu _\theta u^\theta -u^\mu R^\nu_\theta u^\theta )\right]\\
 &+ \left[P_3(u.\nabla)^3 T + P_4(u.\nabla) R +P_5 (u.\nabla)R_{00}\right] u^\mu\\
 &+P^{\mu a}\left[ P_6\nabla_a R_{00}  + P_7 \nabla_a R\right]
\end{split}
\end{equation}
In the first and the third term the two independent $I_3$ type fluid data are chosen 
to be $ (u.\nabla)^2\nabla^\mu T$ and $(u.\nabla)^3 T$.
The independent $I_3$ type curvature data are chosen from the list given in section 
\ref{sec:class}. 
The second term (with the coefficient $P_2$) contains the independent
 vector $P^{\mu a}u^b(u.\nabla) R_{ab}$.

 Here all the terms in the third and the fourth line produce independent $I_4$ 
type scalars at
 fourth order and therefore they are all set to zero. The divergence of the first 
two terms vanish identically.
 
 So finally this part of the entropy current has only two terms and both the terms 
have zero divergence.
 \begin{equation}\label{tinent2}
\begin{split}
&\tilde J^\mu|_{\text{3rd order/$I_3$ type}}\\
 =& \nabla_\nu \left[P_1(u^\nu (u.\nabla) \nabla^\mu T 
- u^\mu (u.\nabla)\nabla^\nu T)\right] 
+ \nabla_\nu \left[P_2(u^\nu R^\mu _\theta u^\theta 
-u^\mu R^\nu_\theta u^\theta )\right]\\
 \end{split}
 \end{equation}

 The part multiplying the $C_{1,2}$ type terms have total 
29 coefficients to begin with. We shall try to write them in a way so that the
 computation becomes simpler.
 In table (\ref{table:scalcombination}) and table (\ref{table:veccombination}) 
we have listed each of the independent $C_{1,2}$ type fluid data and then the 
independent combination 
through which this data 
 has entered the entropy current. In table (\ref{table:C111type}) we have listed 
the relevant $C_{1,1,1}$ type scalars and vectors and also their coefficients in 
the entropy current. In all these cases, to begin with the coefficients are some 
unspecified functions of temperature.

 \begin{table}[ht]
\caption{$C_{1,2}$ type Scalars (Fluid data)} 
\vspace{0.5cm}
\centering 
\begin{tabular}{|c| c|} 
\hline\hline 

\hline
Scalars as listed before&Combination that enters entropy current\\[0.5ex]
\hline
$\Theta (u.\nabla)\Theta$ &$Q_1\left[\Theta (u.\nabla)\Theta\right] u^\mu$
\\[0.5ex]
\hline
$\sigma_{\mu\nu}(u.\nabla)\sigma^{\mu\nu}$ &$Q_2 \left[\sigma_{ab}(u.\nabla)\sigma^{ab}\right]u^\mu$\\[0.5ex]
\hline
${\mathfrak a}_\mu\nabla_a\sigma^{a\mu}$ &$\nabla_\mu\left[Q_3\left(u^\mu\sigma^{a\nu}-u^\nu\sigma^{a\mu}\right){\mathfrak a}_a\right]$
\\[0.5ex]
\hline
${\mathfrak a}_\mu\nabla_a\omega^{a\mu}$&$\nabla_\mu\left[Q_4\left(u^\mu\omega^{a\nu}-u^\nu\omega^{a\mu}\right){\mathfrak a}_a\right]$\\[0.5ex]
\hline
\hline
\end{tabular}
\label{table:scalcombination} 
\end{table}

\begin{table}[ht]
\caption{$C_{1,2}$ type Vectors (Fluid data)} 
\vspace{0.5cm}
\centering 
\begin{tabular}{|c| c|} 
\hline\hline 

\hline
Vectors as listed before&Combination that enters entropy current\\[0.5ex]
\hline
$\Theta \nabla^\mu\Theta $&$\nabla_\nu \left[Q_5~\Theta(u^\mu g^{a\nu} - u^\nu g^{a\mu}){\mathfrak a}_a\right]$
\\[0.5ex]
\hline
$\Theta \nabla_\nu \omega^{\mu\nu}$ &$\nabla_\nu\left[Q_6~\Theta\omega^{\mu\nu}\right]$\\[0.5ex]
\hline
$\omega^{\mu\nu}\nabla_a\sigma^a_\nu$ &$\nabla_\nu
 \left[Q_7\left(\omega^{\mu\theta}\sigma_\theta^\nu
   -\omega^{\nu\theta}\sigma_\theta^\mu\right)\right]$
\\[0.5ex]
\hline
${\mathfrak a}_a(\nabla^\mu\nabla^a T)$&$Q_{8} ~{\mathfrak a}_a(\nabla^\mu\nabla^a T)$\\[0.5ex]
\hline
$\omega ^{ab}\nabla^\mu \omega_{ab}$&$Q_{9}~\omega ^{ab}\nabla^\mu \omega_{ab}$\\[0.5ex]
\hline
${\mathfrak a}^\mu (u.\nabla)\Theta$&$Q_{10}\left[{\mathfrak a}^\mu (u.\nabla)\Theta -u^\mu ({\mathfrak a}.\nabla)\Theta \right]$\\[0.5ex]
\hline
$\omega_{\mu\nu} \nabla^\nu \Theta$&$\omega^{\mu\nu}\nabla_\nu\left(Q_{11}\Theta \right)$\\[0.5ex]
\hline
$\sigma^\mu_\nu\nabla^\nu\Theta$&$Q_{12}~\sigma^\mu_\nu\nabla^\nu\Theta$\\[0.5ex]
\hline
$\sigma^\mu_\nu\nabla_\theta\sigma^{\theta\nu}$&$ Q_{13}~\sigma^\mu_\nu\nabla_\theta\sigma^{\theta\nu}$\\[0.5ex]
\hline
$ P^\mu_c\sigma_{ab}\nabla^{\langle c}\sigma^{ab\rangle}$&$P^\mu_c\sigma_{ab}\nabla^{\langle c}\sigma^{ab\rangle}$\\[0.5ex]
\hline
$P^{\mu c}\sigma^{ab}\left(\nabla _a \omega_{b c}-\frac{P_{ab}}{3}\nabla^k\omega_{kc}\right)$&$Q_{15}~P^{\mu c}\sigma^{ab}\left(\nabla _a \omega_{b c}-\frac{P_{ab}}{3}\nabla^k\omega_{kc}\right)$\\[0.5ex]
\hline
\hline
\end{tabular}
\label{table:veccombination} 
\end{table}

We shall start our analysis by computing the divergence of the 
terms appearing in table (\ref{table:scalcombination}) and (\ref{table:veccombination}).
   \begin{itemize}
      \item The divergence of the terms with coefficients ($Q_i,~~i=3,\cdots,7$)
 vanish identically.
      \item It turns out that the term with coefficient $Q_8$ is the only term 
which produces
 ${\mathfrak a}_\mu$ times a third order $I_3$ type vector. Therefore $Q_8$ has 
to be set to zero.
      
\item    Similarly if we analyse only the third order entropy current, the term 
with coefficient
 $Q_9$ is the only term that produces $l_\mu$ times a third order $I_3$ type 
pseudo-vector in
 the fourth order divergence. However a similar term is also produced when the 
divergence of the 
$B_1$ term in the second order entropy current is computed upto fourth order.
\begin{equation}\label{divB1}
\begin{split}
\nabla_\mu\left[B_1 \omega^2 u^\mu \right] &= B_1 \omega^2\Theta 
+ [(u.\nabla)B_1]\omega^2 + 2 B_1\omega^{ab}(u.\nabla)\omega_{ba}\\
\end{split}
\end{equation}
 
 Since there is no second order on-shell pseudo vector, 
$\omega^{ba}(u.\nabla)\omega_{ba}$
 must contain a third order pseudo-vector times $l_\mu$. 
Then in the final fourth order divergence 
the total coefficient of such term (ie. the term proportional to 
third order pseudo-vector times $l_\mu$) will
 be a linear combination of $B_1 $ and $Q_9$, which should be set to zero.
 
 But to simplify the calculation instead we shall introduce a third order 
shift in the `$B_1$ term' of the
 second order entropy current and will consider the following term 
$\left[B_1 \omega^2 u^\mu -\frac{2B_1}{Ts} \omega^{\mu b}\nabla_a\Pi^a_b\right]$. 
The divergence of the shifted `$B_1$ term' no longer contains the terms 
proportional to third order pseudo-vector times $l_\mu$. 
(The releveant part for the divergence of the shifted 
`$B_1$' term is computed in \eqref{termb1}.
    Once this shift is done, $Q_9$ also has to be set to zero, since 
now this is the only term which 
produces $l_\mu$ times a third order $I_3$ type pseudo-vector.
 \end{itemize}    
    For the rest of the 8 terms we have to compute the divergence explicitly. 
However we are
 interested in those terms in the fourth order divergence which does not have any 
explicit factor of 
$\Theta$ or $\sigma_{\mu\nu}$. This simplifies the calculation.
   For example, in the divergence of the term with coefficient $Q_1$, the only
 contribution which will
 be relevant for our purpose is  $\bigg(Q_1~[(u.\nabla)\Theta]^2\bigg)$.
    
  \begin{itemize}
  \item The relevant part of the divergence of the last two terms with coefficients
 $Q_{14}$
 and
 $Q_{15}$ are the following.
  
  \begin{equation}\label{q14q15}
  \begin{split}
  \nabla_\mu\left[ Q_{14}~P^\mu_ c\sigma_{ab}\nabla^{\langle c}\sigma^{ab\rangle}\right]&\Rightarrow~~ Q_{14}~P^\mu_\nu\left[\nabla_{\langle\mu}\sigma_{ab\rangle}\right]\left
  [\nabla^{\langle\nu}\sigma^{ab\rangle}\right]\\
  \\
  \nabla_\mu\left[~Q_{15}~P^{\mu c}\sigma^{ab}\left(\nabla _a \omega_{b c}-\frac{P_{ab}}{3}\nabla^k\omega_{kc}\right)\right]&\Rightarrow~~ Q_{15}~P^{\nu c}\left[\nabla_\nu \sigma^a_b\right]\left[\nabla _a \omega_{b c}-\frac{P_{ab}}{3}\nabla^k\omega_{kc}\right]
  \end{split}
 \end{equation}
  Here in the first line we get a term proportional to $\text{(spin-3)}^2$ and in
 the second line we get a term proportional to $\text{(pseudo-tensor)}^2$. It will 
turn out that such terms cannot occur in any other place. Positivity of the divergence
 will be satisfied if both $Q_{14}$ and $Q_{15}$ are positive. Therefore these terms 
will not produce any constraint on the second order transport coefficients.

  \item The other five terms where the relevant parts are easy to calculate are the 
following.
 \begin{equation}\label{q12111213}
 \begin{split}
 \nabla_\mu\left[Q_1~u^\mu\Theta (u.\nabla)\Theta \right]&\Rightarrow~~ Q_1[(u.\nabla)\Theta]^2\\
 \nabla_\mu\left[Q_2 ~u^\mu \sigma_{ab}(u.\nabla)\sigma^{ab}\right]&\Rightarrow~~ Q_2[(u.\nabla)\sigma_{ab}][(u.\nabla)\sigma^{ab}]\\
 \\
 \nabla_\mu\left[Q_{12}~\sigma^\mu_\nu\nabla^\nu\Theta\right]
 &\Rightarrow~~ Q_{12}[\nabla_\mu\sigma^\mu_\nu][\nabla^\nu\Theta]\\
 \nabla_\mu\left[Q_{13}~\sigma^\mu_\nu\nabla_\theta\sigma^{\theta\nu}\right]&\Rightarrow~~ Q_{13}\left[\nabla_\mu\sigma^\mu_\nu\right]
 \left[\nabla_\theta\sigma^{\theta\nu}\right]\\
   \end{split}
 \end{equation}
 
 \item The relevant part in the divergence of the terms with coefficients 
$Q_{10}$ and $Q_{11}$ 
are  more complicated. 
 
 The divergence of the `$Q_{11}$-term' is given by the following expression.
   \begin{equation}\label{q11}
 \begin{split}
 &\nabla_\mu\left[\omega^{\mu\nu}\nabla_\nu\left(Q_{11}\Theta \right)\right]\\
= &~Q_{11}\left[\nabla_\mu\omega^{\mu\nu}\right]
 \left[\nabla_\nu\Theta\right] + \left[\nabla_\mu\omega^{\mu\nu}\right] 
\left[\nabla_\nu Q_{11}\right]\Theta\\
 \Rightarrow&-Q_{11}\left[\nabla^\mu\Theta\right]
 \bigg[- P_{\nu\mu}\nabla_a\sigma^{a \nu} + \frac{2}{3}P_{\nu\mu}\nabla^\nu\Theta
  + P_{\mu\nu} u_a R^{a \nu} + {\mathfrak a}_b \omega^{b\mu}\bigg]\\
&-Q_{11}\omega^2(u.\nabla)\Theta\\
  \end{split}
 \end{equation}
 where in the last line we have used the identity \eqref{id4} 
 and ignored the terms proportional to $\Theta$ and $\sigma_{\mu\nu}$.
 
 The divergence of the `$Q_{10}$-term' is given by
 \begin{equation}\label{q10}
 \begin{split}
& \nabla_\mu\bigg(Q_{10}\left[{\mathfrak a}^\mu (u.\nabla)\Theta -u^\mu ({\mathfrak a}.\nabla)\Theta \right]\bigg)\\
=&-T\left(\frac{dQ_{10}}{dT} \right){\mathfrak a}^2 (u.\nabla)\Theta + s\left(\frac{d Q_{10}}{ds}\right) \Theta ({\mathfrak a}.\nabla)\Theta\\
&+Q_{10}\left[(\nabla.{\mathfrak a})(u.\nabla)\Theta + {\mathfrak a}^\mu(\nabla_\mu u^a)(\nabla_a \Theta) -(\nabla_b\Theta) (u.\nabla) {\mathfrak a}^b -\Theta({\mathfrak a}.\nabla)\Theta\right]\\
\Rightarrow&~-\left(T\frac{dQ_{10}}{dT} + Q_{10}\right){\mathfrak a}^2 (u.\nabla)\Theta \\
&+Q_{10}\bigg[\omega^2(u.\nabla)\Theta +[(u.\nabla)\Theta]^2 + R_{00} (u.\nabla)\Theta
 -\left(\frac{s}{T}\frac{dT}{ds}\right)P_{ab}(\nabla^a\Theta)(\nabla^b\Theta)\bigg]
 \end{split}
 \end{equation}
 To express the divergence in the chosen basis of independent data we have used the
 identities \eqref{id1}, \eqref{id2} and \eqref{id3}.
 
 Here also in the final expression we have ignored the terms proportional to $\Theta$ 
and $\sigma_{\mu\nu}$.
   \end{itemize} 
\begin{table}[ht]   
\caption{$C_{1,2}$ type Scalars (Curvature data)} 
\vspace{0.5cm}
\centering 
\begin{tabular}{|c| c|} 
\hline\hline 

\hline
Scalars as listed before&Combination that enters entropy current\\[0.5ex]
\hline
$F_{ab}\sigma^{ab}$ &$P_1 \left(F_{ab}\sigma^{ab}\right)u^\mu$
\\[0.5ex]
\hline
$R_{ab}\sigma^{ab}$ &$P_2 \left(R_{ab}\sigma^{ab}\right)u^\mu$\\[0.5ex]
\hline
$\Theta R$ &$P_3~u^\mu \Theta R$
\\[0.5ex]
\hline
$\Theta R_{00}$&$P_4~u^\mu \Theta R_{00}$\\[0.5ex]
\hline
$u_a{\mathfrak a}_bR^{ab}$&$P_5~u^\mu \left(u_a{\mathfrak a}_bR^{ab}~~\right)$\\[0.5ex]
\hline
\hline
\end{tabular}
\label{table:scalcombcurv} 
\end{table}

\begin{table}[ht]
\caption{$C_{1,2}$ type Vectors (Curvature data)} 
\vspace{0.5cm}
\centering 
\begin{tabular}{|c| c|} 
\hline\hline 

\hline
Vectors as listed before&Combination that enters entropy current\\[0.5ex]
\hline
${\mathfrak a}^\mu R $&$P_6 ~{\mathfrak a}^\mu R$
\\[0.5ex]
\hline
${\mathfrak a}^\mu R_{00}$ &$P_7~{\mathfrak a}^\mu R_{00}$\\[0.5ex]
\hline
$u^a R_{ab}\omega^{b\mu}$ &$P_8~u^a R_{ab}\omega^{b\mu}$
\\[0.5ex]
\hline
${\mathfrak a}_\nu R^{\mu\nu}$&$P_{9} \left(R^{\mu\nu} - \frac{1}{2}g^{\mu\nu} R\right){\mathfrak a}_\nu$\\[0.5ex]
\hline
${\mathfrak a}_\nu F^{\mu\nu}$&$P_{10}\left(F^{\mu\nu}{\mathfrak a}_\nu -R_{00}{\mathfrak a}^\mu +u^\mu u_a {\mathfrak a}_b R^{ab}\right)$\\[0.5ex]
\hline
$u_a R^{a\mu bc}\omega_{bc}$&$P_{11}\left(u_a R^{a\mu\alpha\beta}\omega_{\alpha\beta} + 2\omega^{\mu\alpha} u^a R_{a\alpha}\right)$\\[0.5ex]
\hline
$u_a\Theta R^{a\mu}$&$P_{12} ~u_a\Theta R^{a\mu}$\\[0.5ex]
\hline
$u^a R_{ab}\sigma^{b\mu}$&$P_{13}~u^a R_{ab}\sigma^{b\mu}$\\[0.5ex]
\hline
$u_a R^{ab \mu c}\sigma_{bc}$&$P_{14}~u_a R^{ab \mu c}\sigma_{bc}$\\[0.5ex]
\hline
\hline
\end{tabular}
\label{table:veccombcurv} 
\end{table}

Next we shall compute the divergence of the curvature type data appearing in 
table (\ref{table:scalcombcurv}) and (\ref{table:veccombcurv}).
  \begin{itemize}
    \item It turns out that the divergence of the terms multiplying $P_5$, $P_6$ 
and $P_7$ are the 
only terms which produce the terms of the form ${\mathfrak a}_\mu$ times an independent 
third 
order $I_3$ type curvature vector. Therefore we have to set $P_5$, $P_6$ and $P_7$  to
 zero.
 \item Similarly the divergence of the term multiplying $P_8$ is the only place where
 $l_\mu$ times a 
third order $I_3$ type curvature pseudo-vector is produced.
 Therefore $P_8$ should also be set to zero. 
   \end{itemize}
  The divergence of the remaining 10 terms have to be computed. Here also we shall 
ignore any 
term that are multiplied by an explicit factor of $\Theta$ or $\sigma_{\mu\nu}$.
\begin{itemize}
 \item First we shall determine the relevant part of the divergence of the terms 
with the coefficients $P_i,~~~i= 1,2,3,4~~ \text{and}~~12,13,14$.
 These are easy to calculate
 \begin{equation}\label{p1234}
 \begin{split}
 \nabla_\mu\left[P_1 \left(F_{ab}\sigma^{ab}\right)u^\mu\right]&\Rightarrow~~P_1~ F^{ab}(u.\nabla)\sigma_{ab}\\
 \nabla_\mu\left[P_2 \left(R_{ab}\sigma^{ab}\right)u^\mu\right]&\Rightarrow~~P_2~ R^{ab}(u.\nabla)\sigma_{ab}\\
 \nabla_\mu\left[ P_3~u^\mu \Theta R\right]&\Rightarrow~~P_3~ R (u.\nabla)\Theta\\
 \nabla_\mu\left[ P_4~u^\mu \Theta R_{00}\right]&\Rightarrow~~P_4 ~R_{00} (u.\nabla)\Theta\\
 \\
\nabla_\mu\left[ P_{12} ~P^{\mu}_{ b}u_a\Theta R^{ab}\right]&\Rightarrow~~P_{12}~ u_aR^{ab}P^{\mu}_{ b}\nabla_\mu\Theta\\
\nabla_\mu\left[P_{13}~u^a R_{ab}\sigma^{b\mu}\right]&\Rightarrow~~P_{13}~u^aR_{ab}P_\nu^b
\nabla_\mu\sigma^{\mu \nu}\\
\nabla_\mu\left[P_{14}~u_a P^\mu_\nu R^{ab \nu c}\sigma_{bc}\right]&\Rightarrow~~P_{14}~u_a P^\mu_\nu R^{ab\nu c}\nabla_\mu\sigma_{bc}
 \end{split}
 \end{equation}
 \end{itemize}

Now we shall compute the divergence of the difficult terms multiplying the 
coefficients $P_9$, $P_{10}$ and $P_{11}$ respectively.
We first analyse the situation where in a given basis all the fluid data are locally
 zero upto the required order and only the curvature data are turned on.
For such configurations it will turn out that only these three terms can produce 
non-zero divergence. They are given by the following expressions\footnote{In this computation,
 apart from the explicit curvature there is one more source for producing the curvature terms. 
These are arising because we want to write the final answer for the fluid data in a given basis as
 chosen in section \ref{sec:class}. For example while computing the left hand side of equation
 \eqref{p9curv}, we shall get a term like $R(\nabla.{\mathfrak a})$. However, our basis of 
independent second order fluid data contains a single scalar $(u.\nabla)\Theta$. Therefore
 we have to express $(\nabla.{\mathfrak a})$ in terms
$(u.\nabla)\Theta$ before setting the fluid data to zero. In this process we shall generate a 
curvature term $R_{00}$ as calculated in equation \eqref{id2}. Similar techniques have been 
used to compute the divergence of the other two terms multiplying $P_{10}$ and $P_{11}$.}.
\begin{equation}\label{p9curv}
\begin{split}
&\nabla_\mu\left[P_9~\left(R^{\mu b} - \frac{1}{2}g^{\mu b} R\right){\mathfrak a}_b\right]
=P_9\left[-\frac{R.R_{00}}{2} + R^{ab}F_{ab} \right]
\end{split}
\end{equation}

\begin{equation}\label{p10curv}
\begin{split}
&\nabla_\mu\left[P_{10}\left(F^{\mu\nu}{\mathfrak a}_\nu -R_{00}{\mathfrak a}^\mu + u^\mu u_a {\mathfrak a}_b R^{ab}\right)\right]
=P_{10}\left[F^{ab}F_{ab} - R_{00}^2\right]
\end{split}
\end{equation}

\begin{equation}\label{p11curv}
\begin{split}
&\nabla_\mu\left[P_{11}\left(u_a R^{a\mu\alpha\beta}\omega_{\alpha\beta} + 2\omega^{\mu\alpha} u^a R_{a\alpha}\right)\right]\\
=~&P_{11}\left[2u^a u^b R_{ac}P^{cd}R_{db}
+A^{\mu\nu\lambda}\left(\frac{1}{2} A_{\mu\nu\lambda}+ A_{\lambda\nu\mu}\right)\right]
\end{split}
\end{equation}
where $A^{\mu\nu\lambda} =u_\rho R^{\rho a b c}P^\mu_a P^\nu _b P^\lambda_c$.

Now from these three equations \eqref{p9curv}, \eqref{p10curv} and \eqref{p11curv} 
we can conclude the following.
\begin{itemize}
\item The last term in the RHS of \eqref{p11curv} contains only  the $(pseudo tensor)^2$
 and a $(vector)^2$, but cannot produce any term proportional to $R_{00}^2$ or 
$F_{ab}F^{ab}$.
Therefore to have positivity of the divergence  for all values of $F^2$ and $R_{00}^2$
 we 
must set $P_{10}$ to zero .
\item Once $P_{10}$ is set to zero, there are no terms in the final expressions of 
divergence 
that contain $\left(R_{\langle\mu\nu\rangle}\right)^2$,
 $\left(F_{\langle\mu\nu\rangle}\right)^2$, $R^2$ or 
$R_{00}^2$. 
Therefore in the full divergence the coefficients of all the terms linear in these
 four independent 
data must be zero.
To satisfy this condition we have to set the following coefficients to zero.
 \begin{enumerate}
  \item $P_1 = 0$ as it is the total coefficient of 
the term $F^{ab}(u.\nabla)\sigma_{ab}$.
 \item $P_2 = 0$ as it is the total coefficient of 
the term $R^{ab}(u.\nabla)\sigma_{ab}$.
 \item $P_3 = 0$ as it is the total coefficient of 
the term $R(u.\nabla)\Theta$.
 \item $Q_{10} + P_4 = 0$ as it is the total coefficient 
of the term $R_{00}(u.\nabla)\Theta$.
 \item $P_9 = 0$ as it is the total coefficient of the term $R.R_{00}$.
  \end{enumerate}

   {\bf {$C_{1,1,1}$ type data}}

In the fourth order divergence we are not interested in any terms that are multiplied 
by $\Theta$ or $\sigma_{\mu\nu}$. Therefore in the third order entropy current we did 
not need to consider the $C_{1,1,1}$ type terms which contains more than one factor 
of $\Theta$, $\sigma_{\mu\nu}$ or both.

Here we are listing only those terms which we shall require for our analysis.

\begin{table}[ht]   
\caption{$C_{1,1,1}$ type data (Only the relevant ones)} 
\vspace{0.5cm}
\centering 
\begin{tabular}{|c| c|} 
\hline\hline 

\hline
Scalars&Vectors\\[0.5ex]
\hline
$K_1~u^\mu\omega^2\Theta $ &$K_5~\omega^{\mu a}\sigma_{ab}{\mathfrak a}^b$
\\[0.5ex]
\hline
$K_2~u^\mu{\mathfrak a}^2\Theta$ &$K_6~\sigma^{\mu a}\omega_{ab}{\mathfrak a}^b$\\[0.5ex]
\hline
$K_3~u^\mu\left({\mathfrak a}_b{\mathfrak a}_c\sigma^{bc}\right)$ &$K_7~\omega^{\mu a}{\mathfrak a}_b\Theta$
\\[0.5ex]
\hline
$K_4~u^\mu\left(\omega_{ab}\sigma^b_c\omega^{ca}\right)$&$K_8~\omega^2{\mathfrak a}^\mu$\\[0.5ex]
\hline
&$K_9~{\mathfrak a}^2{\mathfrak a}^\mu$\\[0.5ex]
\hline 
&$K_{10} ~\omega^{\mu a}\omega_{ab}{\mathfrak a}^b$\\[0.5ex]
\hline
\hline
\end{tabular}
\label{table:C111type} 
\end{table}

  Now we shall compute relevant part in the divergence of the each of the relevant 
term. It will
 turn out that analysing the explicit expression of the divergence we can further set
 some of the
 non-zero coefficients to zero.
 \\

   \begin{itemize}
 \item  Here also the relevant parts are easy to calculate for the terms with
 the coefficients $K_i,~~i= 1,2,3,4,5,6~\text{and}~7$ .
 \begin{equation}\label{keasy}
 \begin{split}
 \nabla_\mu \left[K_1~u^\mu\omega^2\Theta\right]&\Rightarrow ~~K_1~\omega^2(u.\nabla)\Theta\\
 \nabla_\mu \left[K_2~u^\mu{\mathfrak a}^2\Theta\right]&\Rightarrow ~~K_2~{\mathfrak a}^2(u.\nabla)\Theta\\
 \nabla_\mu\left[K_3~u^\mu\left({\mathfrak a}_b{\mathfrak a}_c\sigma^{bc}\right)\right]&\Rightarrow~~K_3 ~{\mathfrak a}_b{\mathfrak a}_c(u.\nabla)\sigma^{bc}\\
 \nabla_\mu\left[K_4~u^\mu\left(\omega_{ab}\sigma^b_c\omega^{ca}\right)\right]
 &\Rightarrow~~K_4~\omega_{ab}\left[(u.\nabla)\sigma^b_c\right]\omega^{ca}\\
 \\
 \nabla_\mu\left[K_5~\omega^{\mu a}\sigma_{ab}{\mathfrak a}^b\right]&\Rightarrow~~K_5~\omega^{\mu a}\left[\nabla_\mu\sigma_{ab}\right]{\mathfrak a}^b\\
  \nabla_\mu\left[K_6~\sigma^{\mu a}\omega_{ab}{\mathfrak a}^b\right]&\Rightarrow~~K_6~\left[\nabla_\mu\sigma^{\mu a}\right]\omega_{ab}{\mathfrak a}^b\\
  \nabla_\mu\left[K_7~\omega^{\mu b}{\mathfrak a}^b\Theta\right]&\Rightarrow~~K_7~\left[\omega^{\mu b}{\mathfrak a}^b\nabla_\mu\Theta\right]\\
  \end{split}
 \end{equation}
 \item The divergence of the terms with coefficients $K_8$, $K_9$ and $K_{10}$ are 
complicated. 
These are given by the following expressions.
 
 {\it `$K_8$-term'}
 \begin{equation}\label{k8}
 \begin{split}
& \nabla_\mu\left[K_8 ~\omega^2{\mathfrak a}^\mu\right]\\
=&~\left[({\mathfrak a}.\nabla)K_8\right]\omega^2 +2 K_8~\omega^{\mu\nu}({\mathfrak a}.\nabla)\omega_{\nu\mu} +K_8 ~\omega^2 (\nabla.{\mathfrak a})\\
\Rightarrow &~-T\left(\frac{dK_8}{dT}\right){\mathfrak a}^2\omega^2 +2 K_8~\omega^{\mu\nu}({\mathfrak a}.\nabla)\omega_{\nu\mu}\\
&~+ K_8~\omega^2\left[\omega^2 + (u.\nabla)\Theta + R_{00}\right]
 \end{split}
 \end{equation}
 In the last step we have kept only the relevant terms and used \eqref{id1} and 
\eqref{id2}
 for simplification.
 
{\it $ K_9 $-term}
 \begin{equation}\label{k9}
 \begin{split}
 &\nabla_\mu\left[K_9~{\mathfrak a}^2{\mathfrak a}^\mu\right]\\
 =&~{\mathfrak a}^2({\mathfrak a}.\nabla)K_9 + K_9{\mathfrak a}^2(\nabla.{\mathfrak a}) +2 K_9 ~{\mathfrak a}^\mu{\mathfrak a}^\nu\nabla_\mu{\mathfrak a}_\nu\\
 \Rightarrow &~-\left(T\frac{dK_9}{dT}+K_9\right){\mathfrak a}^4
 + K_9~{\mathfrak a}^2\left[\omega^2 + \frac{5}{3}(u.\nabla)\Theta + R_{00}\right]\\
~ &~+2K_9~{\mathfrak a}^\mu{\mathfrak a}^\nu \left[(u.\nabla)\sigma_{\mu\nu}
 + F_{\mu\nu}+\omega_{\mu a}{\omega^a}_\mu\right]
 \end{split}
 \end{equation}
 In the last step we have used equations \eqref{id1}, \eqref{id2} and \eqref{id6}.
 
 {\it $ K_{10} $-term}
 \begin{equation}\label{k10}
 \begin{split}
 &\nabla_\mu\left[K_{10}~\omega^{\mu a}\omega_{a b}{\mathfrak a}^b\right]\\
= &~ K_{10}~\bigg[(\nabla_\mu\omega^{\mu a})\omega_{ab}{\mathfrak a}^b + \omega^{\mu a}{\mathfrak a}^b(\nabla_\mu\omega_{ab})
+ \omega^{\mu a}\omega_{ab}(\nabla_\mu {\mathfrak a}^b)\bigg]\\
&~+(\nabla_\mu K_{10})\omega^{\mu a}\omega_{a b}{\mathfrak a}^b \\
\Rightarrow&~-\left(T\frac{dK_{10}}{dT} + K_{10}\right)
\left({\mathfrak a}_\mu\omega^{\mu a}\omega_{a b}{\mathfrak a}^b\right) 
+ K_{10}~\omega^{\mu a}{\mathfrak a}^b(\nabla_\mu\omega_{ab}) \\
&~+K_{10}~{\omega^\mu}_a\omega^{a \nu}\left[(u.\nabla)\sigma_{\mu\nu} 
+ F_{\mu\nu}+ \omega_{\mu b}{\omega^b}_\nu\right]\\
&~-K_{10}~\omega_{\mu b}{\mathfrak a}^b\bigg[- \nabla_a\sigma^{a \mu} + \frac{2}{3}\nabla^\mu\Theta
  +  u_a R^{a \mu} + {\mathfrak a}_b \omega^{b\mu}\bigg]\\
 \end{split}
 \end{equation}
In the last step we have used relevant part of equations \eqref{id1}, \eqref{id5} 
and \eqref{id7}. 
 \end{itemize}
 
 Now as explained before, all the terms that are linear in $R_{00}$, $R$, $R_{ij}$ and
 $F_{ij}$ should be set to zero. This will imply the following for the $C_{1,1,1}$ 
part of the entropy current.
 
 \begin{enumerate}
 \item $K_8=0$ as it is the total coefficient of the term $\omega^2 R_{00}$.
 \item $K_9=0$ as it is the total coefficient of the term ${\mathfrak a}. F .{\mathfrak a}$
  \item $K_{10}=0$ as it is the total coefficient of the term $Tr[\omega. F .\omega]$
  
 \end{enumerate}
\item Once $K_8$, $K_9$ and $K_{10}$ are zero there are no terms in the fourth order 
divergence which 
are proportional to $\left[\omega^2\right]^2$, 
$\left[{\mathfrak a}^2\right]^2$, 
${\mathfrak a}^2\omega^2$ or $\left[{\mathfrak a}.\omega\right]^2$. 
This will imply that in the divergence the net coefficients of all the terms, 
linear in ${\mathfrak a}^2$, $\omega^2$, $\omega_{\mu a}\omega^{a\nu}$ 
or ${\mathfrak a}_\mu\omega^{\mu\nu}$ should be zero.

To satisfy this condition we have to set all the $K_i$ from $i=1,\cdots,5$ to zero. 
This will also set $Q_{11}$ to zero. 

$K_6$ and $K_7$ get related to $B_1$ and $B_5$ in the following way 
(see \eqref{rele4} and \eqref{termb1}).
$$\frac{K_6}{\eta} = \frac{K_7}{\zeta} =\frac{1}{s}\left(\frac{dB_5}{dT} 
+ \frac{2B_1}{T} + 2 \frac{dB_1}{dT}\right) $$

Absence of these four fourth order terms mentioned above will also impose some
 constraints on the transport coefficients of second 
order stress-tensor by requiring that the coefficients of the four terms
 $\Theta {\mathfrak a}^2$, $ \Theta \omega^2$, $\omega_{ab}\sigma^b_c\omega^{ca}$ 
and ${\mathfrak a}_a \sigma^{ab}{\mathfrak a}_b$ in the third order divergence should
 vanish.

\end{itemize}

 \section{ Constraints on 2nd order transport coefficients}\label{sec:finalconstraint}
In this section we shall finally analyse how this condition of local entropy production
constrains the second order transport coefficients. In the first part of this section 
we shall derive these constraints. These include the set of five relations among the 15
transport coefficients (as mentioned in the section \ref{sec:intro}) and also two
 inequalities 
involving both the first and second order transport coefficients as well as 
some coefficients 
appearing in the entropy current. 

Then in the next subsection we shall compare our final result 
with the answer presented in 
\cite{Kanitscheider:2009as} and \cite{Romatschke:2009kr}.

\subsection{Derivation of the constraints}
At second order, just from symmetry analysis, the stress tensor will have 15 transport 
coefficients.
  \begin{equation}\label{stress2}
  \begin{split}
\Pi_{\mu\nu} =~&T\bigg[ \tau ~(u.\nabla)\sigma_{\langle\mu\nu\rangle} + \kappa_1 R_{\langle \mu\nu\rangle} + \kappa_2 F_{\langle \mu\nu\rangle} +\lambda_0~ \Theta\sigma_{\mu\nu}\\
&+ \lambda_1~ {\sigma_{\langle \mu}}^a\sigma_{a\nu\rangle}+ \lambda_2~ {\sigma_{\langle \mu}}^a\omega_{a\nu\rangle}+ \lambda_3~ {\omega_{\langle \mu}}^a\omega_{a\nu\rangle} + \lambda_4~{\mathfrak a}_{\langle\mu}{\mathfrak a}_{\nu\rangle}\bigg]\\
&+TP_{\mu\nu}\bigg[\zeta_1(u.\nabla)\Theta + \zeta_2 R + \zeta_3R_{00}
 + \xi_1 \Theta^2 + \xi_2 \sigma^2+ \xi_3 \omega^2 
+\xi_4 {\mathfrak a}^2 \bigg]
  \end{split}
  \end{equation}
As explained before, in the expression of the divergence of the entropy current, 
$\Pi^{\mu\nu}$ 
will always appear 
 contracted with  $\sigma_{\mu\nu}$ and $\Pi$ with $\Theta$. Therefore, in $\Pi^{ab}$ 
all the terms,
which have either $\sigma_{ab}$ or $\Theta$ as factors, will finally generate a
 set of quadratic and higher order 
terms in $\sigma_{ab}$ and $\Theta$. Such 
terms are always 
suppressed in dervative expansion over the second order piece of the divergence
provided the shear and the bulk viscosities are non zero.
 Therefore 
the coefficients multiplying these terms can never be constrained from the condition 
of positivity.
Among the 15 transport coefficients, five ($\lambda_0$, $\lambda_1$, $\lambda_2$, 
$\xi_1$ 
and $\xi_2$) are of such type and therefore are completely unconstrained.
 
It turns out that to maintain the positivity of the divergence, the coefficients $\tau$
and $\zeta_1$ have to satisfy some inequalities. This is because 
at fourth order, the divergence of the entropy current will contain terms 
proportional to $\left[(u.\nabla)\sigma\right]^2$ and $\left[(u.\nabla)\Theta\right]^2$ 
whose coefficients are $Q_2$ and $Q_1$ respectively (see \eqref{q12111213}). These two terms, along with 
four other terms ($\sigma^2$, $\Theta^2 $, $\sigma^{\mu\nu} (u.\nabla)\sigma_{\mu\nu}$
and $\Theta  (u.\nabla)\Theta$, appearing in the second and third order pieces 
of the divergence) together can be made positive definite provided
the transport
 coefficients $\tau$ and $\zeta_1$  satisfy the following inequalities.
  \begin{equation}\label{ineq}
  \begin{split}
 ( \zeta_1 - C_\Theta)^2&\leq 4\zeta Q_1\\
 ( \tau - C_\sigma)^2&\leq 4\eta Q_2\\
  \end{split}
  \end{equation}
 Where $C_\Theta$ and $C_\sigma$ are the coefficients 
of the term $\Theta (u.\nabla)\Theta$ and 
  $\sigma^{ab}(u.\nabla)\sigma_{ab}$ respectively in the divergence of the
 third order entropy current.
  \begin{equation}\label{thetasigma}
  \begin{split}
  C_\Theta &= 2 s \frac{dB_5}{ds} - \frac{2}{3}T\frac{dB_5}{dT}
 + 2 B_2 + 2 B_4 s \left(s - T\frac{ds}{dT}\right)\\
  C_\sigma &= T\frac{dB_5}{dT} + 2 B_3\\
  \end{split}
  \end{equation}
But unlike the inequalities for the first order transport 
coefficients ($\eta\geq0$ 
and $\zeta\geq0$) \eqref{ineq}
 involves several free coefficients appearing in the entropy 
current and hence it does not give any relation within the
 transport coefficients themselves.

Now we shall come to those relations which will give some equalities among the remaining 
eight transport coefficients. 
By computing the divergence of the entropy current upto fourth order we can see 
that there 
are no terms proportional to 
$R^2,~R_{00}^2,~R_{ab}R^{ab},~F_{ab} F^{ab}, {\mathfrak a}^4,~\omega^4$ and 
$ ({\mathfrak a}.\omega)^2$. It will imply that the coefficients of the following 
8 terms in 
the divergence of the entropy current have to be zero.
\begin{enumerate}
\item $C_F\equiv$ Coefficient of the term $\sigma_{ab}F^{ab}$
\item $C_R\equiv$ Coefficient of the term $\sigma_{ab}R^{ab}$
\item $C_{\mathfrak a}\equiv$ Coefficient of the term
 ${\mathfrak a}^a{\mathfrak a}^b\sigma_{ab}$
\item $C_{\omega}\equiv$ Coefficient of the term ${\omega}^{ap}{\omega_p}^b\sigma_{ab}$
\item $Q_F\equiv$ Coefficient of the term $\Theta R_{00}$
\item $Q_R\equiv$ Coefficient of the term $\Theta R$
\item $Q_{\mathfrak a}\equiv$ Coefficient of the term $\Theta{\mathfrak a}^2$
\item $Q_{\omega}\equiv$ Coefficient of the term $\Theta{\omega}^2$
\end{enumerate}

  Solving each of the above eight conditions we can express the
 remaining eight
 transport coefficients in terms of the coefficients 
appearing in the divergence of the second order entropy current.
  
    These are given by the following expressions.
  \begin{equation}\label{fconst}
  \begin{split}
  C_R=0&\Rightarrow~\kappa_1= A_3,~~~~~
  C_F=0\Rightarrow~\kappa_2= T\frac{dB_5}{dT}\\
    C_\omega =0&\Rightarrow \lambda_3 = T \frac{dB_5}{dT} - 4 B_1\\
  C_{\mathfrak a}=0&\Rightarrow \lambda_4=-\left[T^2\frac{d^2B_5}{dT^2}
 + T\frac{dB_5}{dT} + 2B_4 T^2\left(\frac{ds}{dT}\right)^2\right]\\
    Q_R=0 &\Rightarrow \zeta_2 =\frac{1}{2}\left[s\frac{dA_3}{ds} - \frac{A_3}{3}\right]\\
  Q_F=0&\Rightarrow \zeta_3 = s\frac{dA_3}{ds} + \frac{A_3}{3}
-\frac{2T}{3}\frac{dB_5}{dT} - 2 B_4 Ts\frac{ds}{dT}\\
  Q_\omega = 0&\Rightarrow \xi_3 =-2 B_4 Ts \frac{ds}{dT} 
+ T \frac{dB_5}{dT}\left[\frac{s}{T}\frac{dT}{ds} -\frac{2}{3}\right] 
-s \frac{dB_1}{ds}\\
  &~~~~~~~~~~~ +  B_1\left[\frac{2s}{T}\frac{dT}{ds} -\frac{1}{3}\right]\\
  Q_{\mathfrak a}=0&\Rightarrow \xi_4 = T^2s\frac{ds}{dT}\frac{dB_4}{dT}  
+ B_4\left[\frac{T^2}{3}\left(\frac{ds}{dT}\right)^2 + 4 T s \frac{ds}{dT}
 +2 T^2s\frac{d^2s}{dT^2}\right]\\
  &~~~~~~~~~~~ + \frac{2}{3}\left(T\frac{dB_5}{dT} + T^2\frac{d^2B_5}{dT^2}\right)\\
&\text{where}~~~ \frac{dB_5}{dT} = \frac{A_3}{T} + \frac{dA_3}{dT}
  \end{split}
  \end{equation}

From \eqref{fconst} we can see that all these eight coefficients can be 
determined in terms of three independent
 coefficients($A_3$, $B_1$ and $B_4$ ) appearing in the third order entropy current. 
Therefore eleminating the three entropy current coefficients, mentioned above
 finally we shall get
five relations among these eight transport coefficients.

  \begin{equation}\label{relationsf}
  \begin{split}
  \kappa_2 =&~ \kappa_1 + T\frac{d\kappa_1}{dT},~~~~~~~
  \zeta_2 =~ \frac{1}{2}\left[s\frac{d\kappa_1}{ds} - \frac{\kappa_1}{3}\right]\\
  \zeta_3 = &\left(s\frac{d\kappa_1}{ds} + \frac{\kappa_1}{3}\right) 
+ \left(s\frac{d\kappa_2}{ds} 
- \frac{2\kappa_2}{3}\right)+\frac{s}{T}\left(\frac{dT}{ds}\right)\lambda_4\\
    \xi_3=&~\frac{3}{4}\left(\frac{s}{T}\right)\left(\frac{dT}{ds}\right)\left(T\frac{d\kappa_2}{dT} + 2\kappa_2\right) -\frac{3\kappa_2}{4} +\left(\frac{s}{T}\right)\left(\frac{dT}{ds}\right)\lambda_4 \\
  &+\frac{1}{4}\left[s\frac{d\lambda_3}{ds} + \frac{\lambda_3}{3} -2 \left(\frac{s}{T}\right)\left(\frac{dT}{ds}\right)\lambda_3\right]\\
  \xi_4 =&~-\frac{\lambda_4}{6} - \frac{s}{T}\left(\frac{dT}{ds}\right)\left(\lambda_4 + \frac{T}{2}\frac{d\lambda_4}{dT}\right) 
  -T\left(\frac{d\kappa_2}{dT}\right)\left(\frac{3s}{2T}\frac{dT}{ds} - \frac{1}{2}\right) \\
  &- \frac{Ts}{2} \left(\frac{dT}{ds}\right)\left(\frac{d^2\kappa_2}{dT^2}\right)
  \end{split}
  \end{equation}
  \\

\subsection{Comparison with \cite{Kanitscheider:2009as}}

  In \cite{Kanitscheider:2009as} authors have constructed some examples of 
 non-conformal 
fluid
 which can be obtained by dimensional reduction of some higher dimensional 
conformal
 theory. 
The entropy of such non conformal fluid is proportional to $T^{2\sigma -1}$
 where
 $2\sigma$
 was the dimension of the space-time before the reduction. Since this particular
nonconformal fluid
satisfies the condition of `positivity' of the divergence of the entropy current 
by construction
 the transport coefficients should also obey the relations listed in \eqref{relationsf}.
  
  Below we are quoting the values of some transport coefficients for such non-conformal
 fluids. 
These are the transport coefficients which enter the 5 relations in \eqref{relationsf}.
  \begin{equation}\label{skendmatch}
  \begin{split}
    \lambda_3 = \Lambda_3 T^{2\sigma -3},
~~~&\xi_3 =\frac{2\sigma -4}{3(2\sigma-1)}\lambda_3\\
    \kappa_1 = \kappa T^{2\sigma -3},
~~~&\zeta_2 =\frac{2\sigma -4}{3(2\sigma-1)}\kappa_1\\
    \kappa_2 = (2\sigma-2)\kappa_1,
~~~&\zeta_3 =\frac{2\sigma -4}{3(2\sigma-2)}\kappa_2\\
    \lambda_4 =0,~~~&\xi_4 = 0
  \end{split}
  \end{equation}
  where $\Lambda_3$ and $\kappa$ are two dimensionful constants
 which depend on  the 
length of the compactified dimensions but independent of temperature. 
Using the fact
 that 
for such dimensionally reduced nonconformal fluids the entropy can be
 written as
  $$s \propto T^{2\sigma -1}$$
one can check that these values satisfy the relations given in
 \eqref{relationsf}.
 \\

\subsection{Comparison with \cite{Romatschke:2009kr}}

 To compare, first we shall express the eight
relevant transport coefficients (the ones which appear in
equation \eqref{relationsf}) in terms of the 
coefficients as given in \cite{Romatschke:2009kr} . The dictionary is the following.

\begin{equation}\label{comprom}
\begin{split}
T\zeta_2 &= \xi_5^{Rom},~~
T\zeta_3 = \xi_6^{Rom},~~
T\xi_3 = -\xi_3^{Rom}\\
T\lambda_3&=-\lambda_3^{Rom},~~
T\kappa_1 =\kappa^{Rom},~~
T\kappa_2 = 2(\kappa -\kappa^*)^{Rom}
\\
T\xi_4 &= \frac{T^2}{s^2}\left(\frac{ds}{dT}\right)^2\xi_4^{Rom},~~
T\lambda_4=\frac{T^2}{s^2}\left(\frac{ds}{dT}\right)^2\lambda^{Rom}_4 
\end{split}
\end{equation}
The author of \cite{Romatschke:2009kr} has argued for the existence of two
 relations among 5 of these 8 nondissipative transport coefficients. These two relations are not
explicitly presented in the paper \cite{Romatschke:2009kr} in generality.
 However the author of \cite{Romatschke:2009kr}
 appears 
to claim, in the un-numbered equation in section 5 (below equation 32 ) of \cite{Romatschke:2009kr},
that in the special case, when 

\begin{equation}\label{unnu}
T = s^{c_s^2},~~~~\text{and}~~~\kappa^{Rom}\propto\frac{s}{T}
\end{equation} 
the two relations among the five transport coefficients reduce to the folllowing.
\begin{equation}\label{romrel}
\begin{split}
\xi_5^{Rom} &= \frac{\kappa^{Rom}}{3}\left[1 - 3 c_s^2\right]\\
\xi_6^{Rom} + \xi_3^{Rom} &= -\left(\frac{3 c_s^2 -1}{3 c_s^2}\right)\left[\kappa^{Rom}
 + c_s^2\lambda_3^{Rom}\right]
\end{split}
\end{equation} 
The first equation in \eqref{romrel} indeed reduces to the second equation  in \eqref{relationsf}
 when $c_s^2$ is a
 constant. In order to compare our results with the second
of \eqref{romrel}, we subtract the fourth equation from the third equation of \eqref{relationsf} and
then use the first equation of \eqref{relationsf}.

This gives a relationship between all the same transport coefficients 
that appear in the second equation of \eqref{romrel}. However the relationship we find is the following.

\begin{equation}\label{unmatched}
\begin{split}
\zeta_3- \xi_3 =&~ \xi_6^{Rom}+ \xi_3^{Rom}\\
=&\left(s\frac{d\kappa_1}{ds} + \frac{\kappa_1}{3}\right) 
+\frac{1}{4}\left(s\frac{d\kappa_2}{ds} + \frac{\kappa_2}{3}\right) 
-\frac{3}{2}\frac{s}{T}\frac{dT}{ds}\kappa_2\\
&-\frac{1}{4}\left[s\frac{d\lambda_3}{ds} + \frac{\lambda_3}{3} 
-2 \left(\frac{s}{T}\right)\left(\frac{dT}{ds}\right)\lambda_3\right]\\
\text{where}&\\
 \kappa_2 =&~ \kappa_1 + T\frac{d\kappa_1}{dT}
\end{split}
\end{equation}
This relationship does not reduce to the second of \eqref{romrel} after substituing the special case
of \eqref{unnu} with constant $c_s$. 
We do not understand the reason for this disagreement. Perhaps the second of \eqref{romrel} applies 
under more restrictive assumptions than stated
explicitly in \cite{Romatschke:2009kr}. As noted in \cite{Romatschke:2009kr} it certainly applies to the particular 
case, described in \cite{Kanitscheider:2009as}.

  \section{ Conformal limit}
  
  Upto second order in derivative expansion 
the final entropy current (consistent with the 
constraint
of non-negative divergence) is given by the following
 expression
  \begin{equation}\label{duientabar}
\begin{split}
&\tilde J^\mu|_{\text{second order}}\\
 =& \nabla_\nu\left[A_1(u^\mu\nabla^\nu T - u^\nu \nabla^\mu T)\right] + \nabla_\nu \left( A_2 T \omega^{\mu\nu}\right)\\
 & + A_3 \left(R^{\mu\nu} - \frac{1}{2}g^{\mu\nu} R\right) u_\nu
+\left(\frac{A_3}{T} + \frac{dA_3}{dT}\right)\left[\Theta \nabla^\mu T - P^{ab}(\nabla_b u^\mu)( \nabla_a T )\right]\\
 &+(B_1\omega^2 + B_2\Theta^2 + B_3 \sigma^2)u^\mu + B_4\left[(\nabla s)^2
u^\mu + 2 s \Theta \nabla^\mu s\right]\\
\end{split}
\end{equation}

If the theory has conformal symmetry, then the entropy current also should
transform covariantly under a conformal transformation. 
The conformally covariant entropy current is a special case of equation 
\eqref{duientabar}. In this case the only available length scale is provided by
the temperature and therefore the temperature dependence of all the
coefficients are fixed just by dimensional argument and also 
some of the coefficients are related to the others in a way so that the terms that transform 
in-homogeneously under
 conformal transformation cancel.

  At second order in derivative expansion, there are three scalars and two vectors 
\cite{Loganayagam:2008is,Romatschke:2009kr,Bhattacharyya:2008xc} which 
transform covariantly under conformal transformation. In our basis these
 are given by the following combinations
  \begin{equation}\label{congcomb}
  \begin{split}
  {\mathcal S}_1 &=\sigma_{ab}\sigma^{ba},~~~~~~~~~~~~~~~~~~~
    {\mathcal S}_2 = \omega_{ab}\omega^{ba}\\
  {\mathcal S}_3 &= \frac{P^{ab}\nabla_a \nabla_b T}{T} - \frac{P^{ab}(\nabla_a T)(\nabla_b T)}{2 T^2} - \frac{R_{00}}{2}  - \frac{R}{4}  + \frac{\Theta^2}{6}\\
       {\mathcal V}_1^\mu &=P^\nu_a\nabla_\mu \sigma^{\mu a} - 3{\mathfrak a}_\mu \sigma^{\mu\nu} ~~~~~ {\mathcal V}_2^\mu =P^\nu_a\nabla_\mu \omega^{\mu a} - {\mathfrak a}_\mu \omega^{\mu\nu}
  \end{split}
  \end{equation}
  
A conformally covariant entropy current should be expressible only in terms 
of these three scalars and two vectors. So to begin with it can have
 five independent coefficients. Then the constraint of positivity will reduce it to some
special case of \eqref{duientabar}.

Here we have used \eqref{duientabar} to deduce the conformally 
covariant form of the entropy current. First we have fixed the temperature dependence of
the coefficients $A_i$ and $B_i$ by dimensional analysis. Then we have tried to figure 
out the minimal set of relations these coefficients have to satisfy such that all
 the terms 
transforming in-homogeneously under conformal transformation cancel. This means
that one should be able to choose the coefficients $A_i$ and $B_i$ in such a way 
so that the entropy 
current is expressible in terms of these 3 conformal scalars and
   2 conformal vectors. To do this we first rearrange some of the terms appearing in
 equation \eqref{duientabar} assuming that the temperature dependence of  
the coefficients are
 fixed by dimensional analysis. 
   \begin{equation}\label{newarrange1}
   \begin{split}
  & \nabla_\nu\left[A_1(u^\mu\nabla^\nu T - u^\nu \nabla^\mu T)\right]\\
   = ~&A_1 \bigg[u^\mu {\mathcal S}_3 - \frac{1}{2}\left({\mathcal V}_1^\mu + {\mathcal V}_2^\mu\right) + \frac{u^\mu}{2}\left({\mathfrak a}^2 - \Theta^2 + R_{00} + \frac{R}{2}\right)\\
   &~~~~~~-{\mathfrak a}_b\left(\sigma^{b\mu} + \omega^{b\mu}\right) + \frac{\Theta}{3}{\mathfrak a}^\mu - \frac{1}{2} u^k R_{ka}P^{\mu a}\bigg]\\
\end{split}
\end{equation}

\begin{equation}\label{newarrange2}
\begin{split}
  & \nabla_\mu\left[A_2 \omega^{\mu\nu}\right] = A_2\left[ {\mathcal V}_2^\nu -  {\mathcal S}_2 u^\nu\right]\\
  & A_3\left(R^{\mu\nu} - \frac{1}{2}g^{\mu\nu} R\right)u_\nu = A_3\left[-u^\mu \left(\frac{R}{2} + R_{00}\right) + P^{\mu a} R_{ab} u^b\right]\\
\end{split}
\end{equation}

\begin{equation}\label{newarrange3}
\begin{split}
 & \left(\frac{A_3}{T} + \frac{dA_3}{dT}\right)\left[\Theta \nabla^\mu T - P^{ab}(\nabla_b u^\mu)( \nabla_a T )\right] \\
 =~& 2 A_3 \left[\frac{\Theta^2}{3} u^\mu 
- \frac{2\Theta}{3}{\mathfrak a}^\mu + {\mathfrak a}_b\left(\sigma^{b\mu}
 + \omega^{b\mu}\right)\right]\\
\end{split}
\end{equation}

\begin{equation}\label{newarrange4}
\begin{split}
 &B_4\left[(\nabla s)^2
u^\mu + 2 s \Theta \nabla^\mu s\right]
= T^6 B_4\left[\Theta^2 u^\mu + 9\left({\mathfrak a}^2 u^\mu - \frac{2\Theta}{3}{\mathfrak a}^\mu\right)\right]
 \end{split}
\end{equation}
   
  From these expressions we can see how one should choose the coefficients 
$A_i$ and $B_i$ such that 
all the pieces that transform inhomogeneously under conformal transformation 
cancel.
 The coefficients for a conformally covariant entropy current are given by the 
following expressions.
\begin{equation}\label{conformal1}
\begin{split}
A_1(T) = a_1 ,~~&~~A_2(T) = a_2 ,~~~~A_3(T) = \frac{a_1}{2} T\\
B_1(T) =b_1 T, ~~&~~B_2(T)=\frac{2a_1}{9} T,~~~~B_3(T) =b_3 T\\
&~~B_4(T)= -\left(\frac{a_1}{18}\right)T^{-5}
\end{split}
\end{equation}
where all $a_i$ and $b_i$ are constants.
\newline
Therefore the conformally covariant entropy current has four 
independent coefficients ($a_1, ~a_2,~b_1$ 
and $b_3$) when expanded upto second order in derivatives.
When written in terms of these four coefficients the expressions for 
the conformal entropy current is given
 as
\begin{equation}\label{identification}
\begin{split}
&J^\mu_\text{conformal}\\
 &= a_1 T {\cal S}_3 u^\mu +\frac{a_1T}{2}\left({\mathcal V}_1^\mu + {\mathcal V}_2^\mu\right)+a_2 T\left[ {\mathcal V}_2^\nu -  {\mathcal S}_2 u^\nu\right] + b_1 T {\mathcal S}_2 u^\mu + b_2 T {\mathcal S}_1 u^\mu\\
&=T\left[a_1  {\cal S}_3+b_2 {\cal S}_1 + \left(b_1 - a_2\right){\cal S}_2\right]u^\mu
+T \left(a_2+\frac{a_1}{2}\right){\mathcal V}_2^\mu  +\frac{a_1T}{2}{\mathcal V}_1^\mu 
\end{split}
\end{equation}
This expression coincides with the expression presented in   \cite{Bhattacharyya:2008xc} 
and \cite{Romatschke:2009kr} with the following identification.
\begin{equation}\label{identify}
\begin{split}
&a_1T = 4A^{Rom}_3\\
&b_2 T = \frac{A^{Rom}_1}{4} - \frac{A^{Rom}_3}{2} + \frac{B^{Rom}_1}{4}\\
&T(b_1-a_2) = A^{Rom}_2 + 2 A^{Rom}_3 - B^{Rom}_2\\
&\frac{a_1T}{2}=B^{Rom}_1\\
&T \left(a_2+\frac{a_1}{2}\right)=B^{Rom}_2\\
\end{split}
\end{equation}
where $A_i^{Rom}$ and $B_i^{Rom}$ are the coefficients in the conformal entropy 
current as defined in \cite{Romatschke:2009kr}.

Substituting the relations \eqref{conformal1} in \eqref{fconst} one can 
see that in 
conformal case $\zeta_2$, $\zeta_3$, $\xi_3$, $\xi_4$ and $\lambda_4$ vanish and 
$\kappa_2$ is related to $\kappa_1$ as
$$\kappa_2 = 2\kappa_1$$

However once the stress tensor is conformally covariant, all these vanishing of 
the coefficients and the
 relation between $\kappa_1$ and $\kappa_2$ are automatic (If these relations 
were not true then the stress tensor 
would have some terms which will transform in-homogeneously 
under conformal transformation). Therefore we can say that the existence of an
entropy with positive divergence does not constrain the uncharged conformal fluid.

\section{Acknowledgement}
I would like to thank Shiraz Minwalla for suggesting this problem,
collaborating in the initial part of the calculation and providing 
guidance at every stage. I would like to thank T. Sharma
and S. Jain for rechecking the calculation presented in section \ref{sec:class}.
 I would also like to thank N. Banerjee, S. Jain,
 T. Sharma and everyone in HRI for useful discussion. 
Finally I would like to acknowledge 
our debt
to the people of India for their generous and steady support to research in
 the basic science.

\section{Appendices}
\appendix

   \section{Identities}
   Here we list the identities that we have used to calculate the divergence and then
to transform the answer to the required basis.
   \begin{equation}\label{id1}
   \begin{split}
& (u.\nabla)s+s \Theta
 =\frac{\eta\sigma^2 + \zeta\Theta^2}{T}+ \cdots\\
 &P^{\mu\nu}\nabla_\nu T+T {\mathfrak a}^\mu =\frac{P^\mu_a\nabla_\nu\left[\eta\sigma^{\nu a} + \zeta\Theta P^{\nu a}\right]}{s}+\cdots\\ 
 \end{split}
 \end{equation}
 where the RHS of the second equation can be further simplified.
 \begin{equation}\label{id1st}
 \begin{split}
 &\frac{P^\mu_a\nabla_\nu\left[\eta\sigma^{\nu a} + \zeta\Theta P^{\nu a}\right]}{s}\\
 =&~\frac{1}{s}\left[-T\frac{d\eta}{dT}{\mathfrak a}_b\sigma^{b \mu} + \left(\zeta - T\frac{d\zeta}{dT}\right){\mathfrak a}^\mu\Theta + \eta P^\mu_a\nabla_b\sigma^{ab} + \zeta P^{\mu a}\nabla_a\Theta\right]
   \end{split}
   \end{equation}
   
   \begin{equation}\label{id2}
   \begin{split}
   (\nabla.{\mathfrak a}) &= \left(\sigma^2 + \omega^2 + \frac{\Theta^2}{3}\right) + (u.\nabla)\Theta + R_{00} \\
   \end{split}
   \end{equation}
   
   \begin{equation}\label{id3}
   \begin{split}
   (u.\nabla) {\mathfrak a}_\nu &= 
{\mathfrak a}_\nu \Theta\left[2\frac{s}{T}\frac{dT}{ds} 
- \frac{4}{3} - s \frac{d s}{dT}\frac{d^2T}{ds^2}\right]
- {\mathfrak a}^b\sigma_{\nu b} \\
   &~~~+\left[u_\nu {\mathfrak a}^2 - {\mathfrak a}^b\omega_{\nu b} + \frac{s}{T}\frac{dT}{ds}P^\alpha_\nu \nabla_\alpha\Theta\right] + \cdots
   \end{split}
   \end{equation}

\begin{equation}\label{id4}
\begin{split}
P_{\nu\mu}\nabla_a\omega^{a \nu} &=  P_{\nu\mu}\nabla_a\sigma^{a \nu} - \frac{2}{3}P_{\nu\mu}\nabla^\nu\Theta - P_{\mu\nu} u_a R^{a \nu} \\
&~~~- {\mathfrak a}^b\left(\sigma_{b\mu} + \omega_{b\mu}\right) 
\end{split}
\end{equation}   

\begin{equation}\label{id5}
\begin{split}
\omega^{\nu\mu}(u.\nabla)\omega_{\mu\nu} &= -2\omega_{ab}\sigma^b_c\omega^{ca} - \omega^2\left(\frac{2\Theta}{3} + \frac{u.\nabla T}{T}\right)\\
&+\frac{1}{Ts}\left[1+ \frac{T}{s}\frac{ds}{dT}\right]{\mathfrak a}_\mu \omega^{\mu\nu}\left(\nabla_a\Pi^a_\nu\right) 
+\frac{\omega^{\mu\nu}\nabla_\mu\left(\nabla_a\Pi^a_\nu\right)}{Ts} \\
&+\frac{\omega^2}{Ts}\left(\sigma_{\mu\nu}\Pi^{\mu\nu} +\frac{\Theta}{3}\Pi\right)+ \cdots
\end{split}
\end{equation}
 
  \begin{equation}\label{id6}
   \begin{split}
   {\mathfrak a}^\mu{\mathfrak a}^\nu(\nabla_\mu{\mathfrak a}_\nu)
    &= {\mathfrak a}^\mu{\mathfrak a}^\nu \left[\sigma_\mu^a\sigma_{a\nu} 
+ \omega_{\mu a}{\omega^a}_\nu  +\frac{2\Theta}{3}\sigma_{\mu\nu}\right]\\
    &~+{\mathfrak a}^\mu{\mathfrak a}^\nu \left[(u.\nabla)\sigma_{\mu\nu} + F_{\mu\nu} +\frac{P_{\mu\nu}}{3}(u.\nabla)\Theta\right] \\
    &~+ \frac{{\mathfrak a}^2\Theta^2}{9}-\left({\mathfrak a}^2\right)^2
   \end{split}
   \end{equation}

    \begin{equation}\label{id7}
   \begin{split}
   {\omega^\mu}_a\omega^{a \nu}(\nabla_\mu{\mathfrak a}_\nu)
    &= {\omega^\mu}_a\omega^{a \nu} \left[\sigma_\mu^b\sigma_{b\nu} + \omega_{\mu b}{\omega^b}_\nu 
+\frac{2\Theta}{3}\sigma_{\mu\nu}\right]\\
    &~+{\omega^\mu}_a\omega^{a \nu} \left[(u.\nabla)\sigma_{\mu\nu} + F_{\mu\nu} +\frac{P_{\mu\nu}}{3}(u.\nabla)\Theta\right] \\
    &~+ \frac{\omega^2\Theta^2}{9}-{\mathfrak a}_\mu {\mathfrak a}_\nu {\omega^\mu}_a\omega^{a \nu}
   \end{split}
   \end{equation}  
   
 \begin{equation}\label{id10}
   \begin{split}
 \sigma^{\mu\nu}(\nabla_\mu{\mathfrak a}_\nu)
    &= \sigma^{\mu\nu}\left[\sigma_\mu^a\sigma_{a\nu} + {\omega_\mu}^a\omega_{a\nu}
+ F_{\mu\nu} + (u.\nabla)\sigma_{\mu\nu} - {\mathfrak a}_\mu{\mathfrak a}_\nu\right]
+ \frac{2\Theta}{3}\sigma^2
   \end{split}
   \end{equation} 

   \begin{equation}\label{id8}
   \begin{split}
  &2 {\mathfrak a}^\alpha\omega^{\mu\nu}\nabla_\nu\omega_{\mu\alpha}
  =~-\omega^{\mu\nu}({\mathfrak a}.\nabla)\omega_{\nu\mu}
 +\omega^2{\mathfrak a}^2 -{\mathfrak a}_\mu\omega^{\mu\nu}\omega_{\nu\alpha}{\mathfrak a}^\alpha
   \end{split}
   \end{equation}
   The identity \eqref{id8} is derived using the following steps.
   \begin{equation*}\label{stepid8}
   \begin{split}
   &\omega^{\mu\nu}\nabla_\nu\left[\nabla_\mu u_\alpha - \nabla_\alpha u_\mu\right]\\
   =~&\omega^{\mu\nu}\nabla_\nu\bigg(2\omega_{\mu\alpha} - u_\mu{\mathfrak a}_\alpha + u_\alpha{\mathfrak a}_\mu\bigg)\\
   =~&2\omega^{\mu\nu}\nabla_\nu\omega_{\mu\alpha} - \omega^2{\mathfrak a}_\alpha 
   +\omega^{\mu\nu}\left({\mathfrak a}_\mu\nabla_\nu u_\alpha + u_{\alpha}\nabla_\nu{\mathfrak a}_\mu\right)\\
   =~&\omega^{\mu\nu}\bigg(\frac{1}{2}\left[\nabla_\nu,\nabla_\mu\right]u_\alpha - \nabla_\alpha(\nabla_\nu u_\mu )-\left[\nabla_\nu,\nabla_\alpha\right]u_\mu\bigg)\\
   =~&\omega^{\mu\nu}\bigg(-\nabla_\alpha\omega_{\nu\mu} + {\mathfrak a}_\mu\omega^{\mu\nu}\nabla_\alpha u_\nu + u^\rho\left[\frac{1}{2}R_{\rho\alpha\mu\nu} - R_{\rho\mu\alpha\nu}\right]\bigg)\\
      =~&\omega^{\mu\nu}\bigg(-\nabla_\alpha\omega_{\nu\mu} + {\mathfrak a}_\mu\omega^{\mu\nu}\nabla_\alpha u_\nu -\frac{ u^\rho}{2}\left[R_{\rho\nu\alpha\mu} + R_{\rho\mu\alpha\nu}\right]\bigg)\\
      =~&\omega^{\mu\nu}\bigg(-\nabla_\alpha\omega_{\nu\mu} + {\mathfrak a}_\mu\omega^{\mu\nu}\nabla_\alpha u_\nu \bigg)\\
      \end{split}
   \end{equation*}
   
   \begin{equation}\label{id9}
   \begin{split}
  2 A^{\mu\nu\lambda}\nabla_\nu\sigma_{\lambda\mu}
  =~&A^{\mu\nu\lambda}\bigg[\nabla_\mu\omega_{\nu\lambda} + \frac{1}{2}A_{\mu\lambda\nu} + A_{\lambda\nu\mu}\\
  &~~~~~~~~ -  \omega_{\nu\lambda}{\mathfrak a}_\nu -2 \omega_{\nu\mu}{\mathfrak a}_\lambda - \frac{2}{3}\left(u_\rho R^{\rho b}P^\nu_b\right)({\mathfrak a}_\nu\Theta)\bigg]
   \end{split}
   \end{equation}
where $A^{\mu\nu\lambda} = u_\rho R^{\rho abc} P^\mu _a P^\nu _b P^\lambda_c$  

   The identity \eqref{id9} can be derived using the similar tricks as in
the identity \eqref{id8}.
\begin{equation}\label{stepid9}
\begin{split}
&A^{\mu\nu\lambda}\nabla_\nu\left[\nabla_\lambda u_\mu + \nabla_\mu u_\lambda\right]\\
=~&A^{\mu\nu\lambda}\nabla_\nu\left[2\sigma_{\lambda\mu} +\frac{ 2}{3} P_{\lambda\mu}\Theta- u_\lambda{\mathfrak a}_\mu - u_\mu{\mathfrak a}_\lambda\right]\\
=~&A^{\mu\nu\lambda}\left(\frac{1}{2}\left[\nabla_\nu,\nabla_\lambda\right]u_\mu + \left[\nabla_\nu,\nabla_\mu\right] u_\lambda +\nabla_\mu (\nabla_\nu u_\lambda)\right)\\
=~&A^{\mu\nu\lambda}\left[\frac{1}{2}A_{\mu\lambda\nu} + A_{\lambda\mu\nu} + \nabla_\mu\omega_{\nu\lambda} -{\mathfrak a}_\lambda\nabla_\nu u_\mu\right]
\end{split}
\end{equation} 
   
 \section{Computation of the divergence}
Here we shall calculate the divergence of the different terms appearing 
in the second order entropy current.
The final expression for the second order entropy current is given in \eqref{duientabar}. 
Here we are quoting the equation again.
    \begin{equation}\label{duientabar2}
\begin{split}
&\tilde J^\mu|_{\text{second order}}\\
 =& \nabla_\nu\left[A_1(u^\mu\nabla^\nu T - u^\nu \nabla^\mu T)\right] + \nabla_\nu \left( A_2 T \omega^{\mu\nu}\right)\\
 & + A_3 \left(R^{\mu\nu} - \frac{1}{2}g^{\mu\nu} R\right) u_\nu
+\left(\frac{A_3}{T} + \frac{dA_3}{dT}\right)\left[\Theta \nabla^\mu T - P^{ab}(\nabla_b u^\mu)( \nabla_a T )\right]\\
 &+(B_1\omega^2 + B_2\Theta^2 + B_3 \sigma^2)u^\mu + B_4\left[(\nabla s)^2
u^\mu + 2 s \Theta \nabla^\mu s\right]\\
\end{split}
\end{equation}
The first two terms (with coefficients $A_1$ and $A_2$ respectively)
 have zero divergence. 
Below we shall calculate the divergence of the rest of the terms. 
As explained before, to determine the constraints of `positivity' we need to calculate the divergence upto fourth order in derivative expansion. However in the fourth order piece of the divergence we need to retain only those terms which do not involve any factor of $\sigma_{\mu\nu}$ or $\Theta$.

{\bf Divergence of the term with coefficient $A_3$:}
\begin{equation}\label{a3term}
\begin{split}
&\nabla_\mu\left[A_3 \left(R^{\mu\nu} - \frac{1}{2}g^{\mu\nu} R\right) u_\nu\right]_\text{upto 3rd order}\\
=~&(\nabla_\mu A_3)\left(R^{\mu\nu} - \frac{1}{2}g^{\mu\nu} R\right) u_\nu
+ A_3\left(R^{\mu\nu} - \frac{1}{2}g^{\mu\nu} R\right)(\nabla_\mu u_\nu)\\
=~&(\nabla_\mu A_3)R^{\mu\nu}u_\nu - \frac{R}{2} (u.\nabla)A_3
+ A_3\left(R^{\mu\nu} - \frac{R}{2}g^{\mu\nu} \right)\left(\sigma_{\mu\nu} -u_\mu {\mathfrak a}_\nu+ P_{\mu\nu}\frac{\Theta}{3}\right)\\
=~&\Theta\left[R_{00}\left(s\frac{dA_3}{ds} + \frac{A_3}{3}\right) + \frac{R}{2}\left(s\frac{dA_3}{ds} - \frac{A_3}{3}\right)\right]\\
& - \left(A_3 + T\frac{dA_3}{dT}\right) (u^\mu {\mathfrak a}^\nu R_{\mu\nu})
 + A_3 R_{\mu\nu}\sigma^{\mu\nu}\\
\end{split}
\end{equation}
\begin{equation*}
\begin{split}
&\nabla_\mu\left[A_3 \left(R^{\mu\nu} 
- \frac{1}{2}g^{\mu\nu} R\right) u_\nu\right]_\text{relevant terms at 4th order}\\
=~&u_\nu P_\mu^a R^{\mu\nu}\left(\frac{1}{s}\frac{dA_3}{dT}\right)
\left[\eta \nabla_b \sigma^{a b} + \zeta \nabla_a \Theta\right]
\end{split}
\end{equation*}
\\

{\bf Divergence of the term with coefficient $\left(\frac{A_3}{T} + \frac{dA_3}{dT}\right)$:}
\begin{equation}\label{b5term}
\begin{split}
&\nabla_\mu\left[\Theta \nabla^\mu B_5 - P^{ab}(\nabla_b u^\mu)( \nabla_a B_5 )\right]_\text{upto 3rd order}\\
=~&s\frac{dB_5}{ds}\Theta\left[ (u.\nabla)\Theta +\sigma^2 +\omega^2 + \frac{\Theta^2}{3}\right]+ T\frac{dB_5}{dT} \left[{\mathfrak a}_\mu u_\nu R^{\mu\nu} + {\mathfrak a}_\mu{\mathfrak a}_\nu \sigma^{\mu\nu} + \frac{{\mathfrak a}^2\Theta}{3}\right]\\
&-\sigma^{\mu\nu}\nabla_\mu\nabla_\nu B_5 
+ \frac{2\Theta}{3}\nabla^2 B_5 -\frac{\Theta}{3}(u.\nabla)^2B_5
\end{split}
\end{equation}
where $$\frac{dB_5}{dT} =\frac{A_3}{T} + \frac{dA_3}{dT}$$
 The three terms in the second line of equation \eqref{b5term} can be simplified further.

\begin{equation*}
\begin{split}
&(u.\nabla)^2B_5= \left(s^2\frac{d^2B_5}{ds^2} + s\frac{dB_5}{ds}\right) \Theta^2 - s\frac{dB_5}{ds}(u.\nabla)\Theta + T \frac{dB_5}{dT}{\mathfrak a}^2\\
\\
&\nabla^2 B_5 \\
&= -\Theta^2\left[s^2\frac{d^2B_5}{ds^2} + \frac{T}{3}\frac{dB_5}{dT}\right] + \left[s\frac{dB_5}{ds} - T \frac{dB_5}{dT}\right](u.\nabla)\Theta\\
&~+\left(T\frac{dB_5}{dT} + T^2\frac{d^2B_5}{dT^2}\right){\mathfrak a}^2 - T\frac{dB_5}{dT}(\sigma^2 + \omega^2 + R_{00})\\
\\
&\sigma^{\mu\nu}\nabla_\mu\nabla_\nu B_5\\
&=\left(2T\frac{dB_5}{dT} +
 T^2\frac{d^2B_5}{dT^2}\right)\left({\mathfrak a}_\mu \sigma^{\mu\nu}{\mathfrak a}_\nu\right) 
+ \left(s\frac{dB_5}{ds}- \frac{2T}{3}\frac{dB_5}{dT}\right)\sigma^2\Theta \\
&~~- T\frac{dB_5}{dT}\sigma_{\mu\nu}\left[ F^{\mu\nu} + (u.\nabla)\sigma^{\mu\nu}
+ \sigma^{\mu a}\sigma_a^\nu 
+ \omega^{\mu a}\omega_a^\nu\right]
\end{split}
\end{equation*}

 The relevant part of the fourth order piece in the divergence is given by the following expression.
 \begin{equation}\label{rele4}
 \begin{split}
 &\nabla_\mu\left[\Theta \nabla^\mu B_5 - P^{ab}(\nabla_b u^\mu)( \nabla_a B_5 )\right]_\text{relevant part at 4th order}\\
 =~& -\frac{dB_5}{dT}\left({\mathfrak a}_\mu\omega^{\mu a} +u_\nu P_{\mu a} R^{\mu\nu}\right)\left(\frac{\eta P^a_b\nabla_\mu \sigma^{\mu b} + \zeta P^{\mu a}\nabla_\mu \Theta}{s}\right)\\
 \end{split}
 \end{equation}
\newpage

{\bf Divergence of the term with coefficient $B_1$:}

\begin{equation}\label{termb1}
\begin{split}
&\nabla_\mu\left[B_1 \omega^2 u^\mu \right]_\text{upto 3rd order}\\
= ~&B_1 \omega^2\Theta + [(u.\nabla)B_1]\omega^2 -2B_1\left[\omega_{ab}\sigma^b_c\omega^{ca} +\omega^2\frac{\Theta}{3} - \frac{s}{2T}\frac{dT}{ds}\omega^2\Theta\right] \\
=~&\left[-s\frac{dB_1}{ds} 
- \frac{B_1}{3} + 2 B_1\left(\frac{s}{T}\frac{dT}{ds}\right)\right]\omega^2\Theta 
- 4 B_1 \sigma_\mu^a\omega^{\mu\nu}\omega_{\nu a}\\
\\
&\nabla_\mu\left[B_1 \omega^2 u^\mu  +\left( \frac{2 B_1}{Ts}\right) \omega^{b\mu}\left(\nabla_a\Pi^a_b\right)\right]_\text{relevant part at 4th order}\\
=~&\left[\frac{2}{s} \left(\frac{dB_1}{dT}\right){\mathfrak a}^\mu \omega _{\mu a}+\left(\frac{2B_1}{Ts}\right)\nabla_\mu \omega^{\mu a}\right]\left(\frac{\eta P^a_b\nabla_c \sigma^{c b} + \zeta P^{c a}\nabla_c \Theta}{s}\right)\\
=~&\left[\frac{2}{s} \left(\frac{dB_1}{dT}\right) + \frac{2B_1}{Ts}\right]{\mathfrak a}^\mu \omega _{\mu a}\left(\frac{\eta P^a_b\nabla_c \sigma^{c b} + \zeta P^{c a}\nabla_c \Theta}{s}\right)\\
&-\frac{2B_1}{Ts}\left[P_{a\nu}\nabla_\mu\sigma^{\mu \nu} -\frac{2}{3}P_a^\mu\nabla_\mu\Theta + R_{\mu\nu} u^\nu P^\mu_a\right]\left(\frac{\eta P^a_b\nabla_c \sigma^{c b} + \zeta P^{c a}\nabla_c \Theta}{s}\right)
\end{split}
\end{equation}

{\bf Divergence of the terms with coefficients $B_2$ and $B_3$:}
\begin{equation}\label{b2b3term}
\begin{split}
&\nabla_\mu\left[B_2 \Theta^2 u^\mu\right]_\text{upto 3rd order}= \Theta^3\left(B_2 - s\frac{dB_2}{ds}\right)+2B_2\Theta(u.\nabla)\Theta\\
&\nabla_\mu\left[B_3 \sigma^2 u^\mu\right]_\text{upto 3rd order} = \sigma^2\Theta\left(B_2 - s\frac{dB_2}{ds}\right)+2 B_3 \sigma_{\mu\nu}(u.\nabla)\sigma^{\mu\nu}
\end{split}
\end{equation}

{\bf Divergence of the term with coefficient $B_4$:}
\begin{equation}\label{b4term}
\begin{split}
& ~~\nabla_\mu\left(B_4\left[(\nabla s)^2
u^\mu + 2 s \Theta \nabla^\mu s\right]\right)|_\text{upto 3rd order}\\
&=-\left(s\frac{dB_4}{ds}+ B_4\right)s^2\Theta^3 -2B_4T^2 \left(\frac{ds}{dT}\right)^2{\mathfrak a}_\mu{\mathfrak a}_\nu\sigma^{\mu\nu}+ 2B_4 s \Theta\nabla^2s\\
&~~ + {\mathfrak a}^2\Theta\left[sT^2\left(\frac{ds}{dT}\right)\left(\frac{dB_4}{dT}\right) + \frac{T^2}{3}\left(\frac{ds}{dT}\right)^2B_4 +2B_4Ts\left(\frac{ds}{dT}\right)\right]
\end{split}
\end{equation}
where $\nabla^2s$ can be further simplified.
\begin{equation*}
\begin{split}
\nabla^2s
 &=-T\left(\frac{ds}{dT}\right)\frac{\Theta^2}{3} + \left(s - T\frac{ds}{dT}\right)(u.\nabla)\Theta \\
 &+ \left[T\frac{ds}{dT}
 + T^2\frac{d^2s}{dT^2}\right]{\mathfrak a}^2 - T\frac{ds}{dT}\left[\sigma^2 + \omega^2 + R_{00}\right]
\end{split}
\end{equation*}

There is no relevant part in the fourth order corrections to 
the equations \eqref{b2b3term} and \eqref{b4term} (i.e. all the terms 
appearing in the fourth order corrections to these equations involve 
atleast one factor of $\Theta$ or $\sigma_{\mu\nu}$).

\bibliographystyle{JHEP}
\bibliography{entropybib}
\end{document}